\documentclass[11pt,preprint]{aastex}

\newcommand{\kms}{~km~s$^{-1}$} 
\newcommand{\teff}{$T_{\rm eff}$}
\newcommand{\logg}{$\log g$}

\newcommand{\vt}{$v_{\rm micro}$}

\newcommand*{\feh}{[Fe/H]}

\newcommand*{\alp}{$\alpha$}

\slugcomment{}

\shorttitle{Extremely metal-poor stars from SDSS/SEGUE}
\shortauthors{Aoki et al.}

\begin{document}



\title{High-Resolution Spectroscopy of Extremely Metal-Poor Stars from
  SDSS/SEGUE: I. Atmospheric Parameters and Chemical Compositions}



\author{Wako Aoki\altaffilmark{1}}

\affil{National Astronomical Observatory, Mitaka, Tokyo
181-8588, Japan}
\email{aoki.wako@nao.ac.jp}
\altaffiltext{1}{Department of Astronomical Science, School of Physical Sciences, The Graduate University of Advanced Studies (SOKENDAI), 2-21-1 Osawa, Mitaka,
Tokyo 181-8588, Japan}

\author{Timothy C. Beers\altaffilmark{2}}
\affil{National Optical Astronomy Observatory, Tucson, AZ 85719, USA}
\email{beers@noao.edu}
\altaffiltext{2}{Department of Physics \& Astronomy and JINA: Joint Institute for
Nuclear Astrophysics, Michigan State University, East Lansing, MI  48824, USA}

\author{Young Sun Lee\altaffilmark{3}}
\affil{Department of Physics \& Astronomy and JINA: Joint Institute for
Nuclear Astrophysics, Michigan State University, East Lansing, MI  48824, USA}
\email{lee@pa.msu.edu}
\altaffiltext{3}{present address: Department of Astronomy, New Mexico State University, Las Cruces, NM 88 88003, USA}

\author{Satoshi Honda}
\affil{Kwasan Observatory, Kyoto University, Ohmine-cho Kita Kazan, Yamashin
a-ku, Kyoto 607-8471, Japan}
\email{honda@kwasan.kyoto-u.ac.jp}

\author{Hiroko Ito}
\affil{Department of Astronomical Science, School of Physical Sciences, The Graduate University of Advanced Studies (SOKENDAI), 2-21-1 Osawa, Mitaka,
Tokyo 181-8588, Japan}

\author{Masahide Takada-Hidai}
\affil{Liberal Arts Education Center, Tokai University, 4-1-1 Kitakaname, Hiratsuka, Kanagawa 259-1292, Japan}
\email{hidai@apus.rh.u-tokai.ac.jp}

\author{Anna Frebel}
\affil{Massachusetts Institute of Technology, Kavli Institute for
Astrophysics and Space Research, 77 Massachusetts Avenue,
Cambridge, MA 02139, USA}
\email{afrebel@mit.edu}

\author{Takuma Suda}
\affil{National Astronomical Observatory, Mitaka, Tokyo
181-8588, Japan}
\email{takuma.suda@nao.ac.jp}

\author{Masayuki Y. Fujimoto}
\affil{Department of Cosmosciences, Graduate School of Science,
Hokkaido University, Kita 10 Nishi 8, Kita-ku, Sapporo 060-0810, Japan}
\email{fujimoto@astro1.sci.hokudai.ac.jp}


\author{Daniela Carollo\altaffilmark{4,5}}
\affil{Macquarie University Research Centre in Astronomy, Astrophysics \&
Astrophotonics}
\altaffiltext{4}{Department of Physics \& Astronomy, Macquarie University, NSW 2109 Australia}
\altaffiltext{5}{INAF-Osservatorio Astronomico di Torino, Strada Osservatorio 20, Pino Torinese, 10020, Torino, Italy}
\email{daniela.carollo@mq.edu.au}

\author{Thirupathi Sivarani} 
\affil{Indian Institute of Astrophysics, 2nd block Koramangala, Bangalore 560034, India}
\email{sivarani@iiap.res.in}

\begin{abstract}

Chemical compositions are determined based on high-resolution
spectroscopy for 137 candidate extremely metal-poor (EMP) stars
selected from the Sloan Digital Sky Survey (SDSS) and its first
stellar extension, the Sloan Extension for Galactic Understanding and
Exploration (SEGUE).  High-resolution spectra with moderate
signal-to-noise (S/N) ratios were obtained with the High Dispersion
Spectrograph of the Subaru Telescope. Most of the sample
(approximately 80\%) are main-sequence turn-off stars, including
dwarfs and subgiants. Four cool main-sequence stars, the most
metal-deficient such stars known, are included in the remaining
sample. Good agreement is found between effective temperatures
estimated by the SEGUE stellar parameter pipeline, based on the
SDSS/SEGUE medium-resolution spectra, and those estimated from the 
broadband $(V-K)_0$ and $(g-r)_0$ colors. Our abundance measurements
reveal that 70 stars in our sample have [Fe/H] $ < -3$, adding a
significant number of EMP stars to the currently known sample. Our
analyses determine the abundances of eight elements (C, Na, Mg, Ca,
Ti, Cr, Sr, and Ba) in addition to Fe. The fraction of carbon-enhanced
metal-poor stars ([C/Fe]$> +0.7$) among the 25 giants in our
sample is as high as 36\%, while only a lower limit on the fraction
(9\%) is estimated for turn-off stars. This paper is the first of a
series of papers based on these observational results. The following
papers in this series will discuss the higher-resolution and
higher-S/N observations of a subset of this sample, the metallicity
distribution function, binarity, and correlations between the chemical
composition and kinematics of extremely metal-poor stars.


\end{abstract}


\section{Introduction}\label{sec:intro}

The formation and evolution of the first generations of stars, once an
entirely theoretical enterprise, has in recent years begun to enter
the realm where observations are placing more and firmer constraints
on the subject.  Pertinent observations range from cosmology to star
formation, stellar evolution, supernova explosions, and early galaxy
formation \citep[e.g., ][]{bromm04, ciardi05}. Surveys for very high
redshift galaxies, QSOs, and gamma-ray bursters have detected objects
at $z\gtrsim 6$, when the age of Universe was only several hundred
million years.  The recently reported high redshift ($z = 2.3$),
extremely metal-poor Damped Lyman-$\alpha$ system by Cooke et
al. (2011; [Fe/H] $\sim -3$) exhibits enhanced carbon ([C/Fe] $=
+1.5$) and other elemental abundance signatures that
\citet{kobayashi11} associate with production by faint supernovae
in the early Universe.

Such studies are complemented by investigations of ancient (but still
shining) stars of the Milky Way and Local Group. The elemental
abundances of the chemically most primitive stars are believed to
record the nucleosynthesis yields of the first generations of objects,
thereby constraining their mass distribution, evolution, and nature of
their supernova explosions (Beers \& Christlieb 2005; Frebel \& Norris
2011). If low-mass ($< 0.8$ M$_{\odot}$) stars were able to form from
primordial, metal-free gas clouds, stars with zero metallicity are
expected to be found in the present Galaxy.

A number of extensive searches for very metal-poor (VMP; [Fe/H]$ <
-2$) and extremely metal-poor (EMP; [Fe/H]$ < -3$) stars in the Galaxy
have been undertaken in the past few decades. Since the discovery of
CD$-$38$^{\circ}$245 with [Fe/H]$\sim -4$ \citep{bessell84}, several
objects having similar metallicity have been found by the HK survey
\citep{beers85,beers92} and studied with follow-up high-resolution
spectroscopy. Stars with even lower metallicities have been found in
recent years, including the ultra metal-poor (UMP; [Fe/H] $< -4$) star
HE~0557-4840 \citep{norris07} and the hyper metal-poor (HMP; [Fe/H] $<
-5$) stars HE~0107-5240 \citep{christlieb02} and HE~1327-2326
\citep{frebel05, aoki06}, based on follow-up observations of
candidates from the Hamburg/ESO Survey \citep[HES; ][]{christlieb03,
  christlieb08}, which has a fainter limiting magnitude and larger
survey volume than the HK survey. Quite recently, a new UMP star with
[Fe/H]$\sim -5$ was discovered by \citet{caffau11} among the candidate
metal-poor stars identified with medium-resolution spectroscopy from
the Sloan Digital Sky Survey (SDSS, see below).



The majority of VMP stars found by the HK survey and the HES,
including the two stars with [Fe/H] $ < -5$, are fainter than
$V\sim13$. Detailed abundance measurements, based on high-resolution
spectroscopy for such stars, has only become possible through the use
of 8-10m class telescopes such as Keck, VLT, and Subaru. Previous
studies of large samples of candidate metal-poor stars from these
surveys have revealed the chemical compositions of stars with
[Fe/H]$\sim -3$ \citep{cayrel04, cohen04, honda04, barklem05, aoki05,
  lai08, bonifacio09}. However, the sample size of stars having even
lower metallicity, in particular for the intrinsically fainter
main-sequence turn-off stars, is still rather small, and the
relationship between the abundance patterns observed for the EMP, UMP,
and HMP stars remains unclear. A large sample of candidate
  metal-poor stars have been provided by SDSS (see below), and
  abundance studies for them based on high-resolution spectroscopy
  have been rapidly growing \citep[e.g., ][]{aoki08, caffau11,
    bonifacio12}. \footnote{After our work is completed, a series of
    papers on a large sample of metal-poor stars by
    \citet{norris12} and \citet{yong12} have appeared. Their sample
    includes some EMP stars discovered by SDSS.}


In this paper, the first of a series, we report on follow-up high-resolution
``snapshot'' ($R \sim 36,000$, $30 \lesssim S/N \lesssim 60$) spectroscopic
observations of a large sample (137) of candidate EMP stars selected from the
SDSS \citep{york00}, and the Sloan Extension for
Galactic Understanding and Exploration (SEGUE) sub-survey of the SDSS
\citep{yanny09}. In this paper we describe the selection of our targets (Section 2), 
the observational and reduction/analysis procedures used (Section 3),
and the determinations of stellar atmospheric parameters and estimates of a limited
number of important elements (C, Na, Mg, Ca, Ti, Cr, Sr, and Ba; Section
4). In Section 4, we also comment briefly on a number of the double-lined (and
one triple lined!) spectroscopic binaries discovered during the course of this
work. In Section 5, we discuss the nature of the carbon-enhanced metal-poor
(CEMP) stars found in our our sample, and the trends and outliers found among
the \alp-elements and the neutron-capture elements for stars in our sample.

Papers to follow in this series will discuss constraints on the
low-metallicity tail of the halo-system metallicity distribution
function, the binarity properties of the sample, and correlations
between the chemical compositions and kinematics of VMP and EMP stars.
Results of higher-$S/N$, higher resolution spectroscopy of a number of
the most interesting stars found during this effort will also be
presented, including the Li abundances for main-sequence turn-off
stars.

\section{Sample Selection}\label{sec:sample}

\subsection{Selection of Candidate EMP Stars from SDSS/SEGUE} \label{sec:sample1}

SDSS-I was an imaging and spectroscopic survey that began routine operations in
April 2000, and continued through June 2005. The SDSS, and its extensions, use a
dedicated 2.5m telescope \citep{gunn06} located at the Apache Point Observatory
in New Mexico. The telescope is equipped with an imaging camera and a pair of
spectrographs, each of which are capable of simultaneously collecting 320
medium-resolution ($R \sim 1800$) spectra over its seven square degree field of
view, so that on the order of 600 individual target spectra and roughly 40
calibration-star and sky spectra are obtained on a given spectroscopic
``plug-plate''. It is important to recall that SDSS imaging
(done in drift-scan mode) has an effective bright limit corresponding to roughly
$g \sim 14.0 - 14.5$, which means that high-resolution spectroscopic follow-up
observations for large samples of these stars is challenging to obtain with
telescopes of 4m aperture and smaller.


The SEGUE sub-survey, carried out as part of SDSS-II, ran from July 2005 to June
2008. SEGUE obtained some 240,000 medium-resolution spectra of stars in the
Galaxy, selected to explore the nature of stellar populations from 0.5 kpc to
100 kpc \citep{yanny09}. These stars, as well as all previous SDSS stellar
observations, were released as part of DR7 \citep{abazajian09}.

The SEGUE Stellar Parameter Pipeline (SSPP) processes the wavelength- and
flux-calibrated spectra generated by the standard SDSS spectroscopic reduction
pipeline, obtains equivalent widths and/or line indices for about 80 atomic or
molecular absorption lines, and estimates the effective temperature, {\teff},
surface gravity, {\logg}, and metallicity, [Fe/H], for a given star through the
application of a number of approaches. A given method is usually optimal over
specific ranges of color and $S/N$ ratio. The SSPP employs 8 primary methods for
the estimation of {\teff}, 10 for the estimation of {\logg}, and 12 for the
estimation of [Fe/H]. The final estimates of the atmospheric parameters are
obtained by robust averages of the methods that are expected to perform well for
the color and $S/N$ obtained for each star. The use of multiple methods allows
for empirical determinations of the internal errors for each parameter, based on
the range of reported values from each method -- typical internal errors for
stars in the temperature range that applies to the calibration stars are
$\sigma$({\teff}) $\sim$ 100 K to $\sim$ 125 K, $\sigma$({\logg}) $\sim $ 0.25
dex, and $\sigma$([Fe/H]) $\sim$ 0.20 dex. The external errors in these
determinations are of a similar size. See \citet{lee08a, lee08b}, 
\citet{allendeprieto08}, \citet{smolinski11}, and \citet{lee11} for
additional discussion of the SSPP.  The SSPP estimates of {\teff},
{\logg}, and {\feh} were also released as part of DR7.

\subsection{Sample Selection for High-Resolution Spectroscopy}\label{sec:sample2}

\begin{figure}
\epsscale{.65}
\plotone{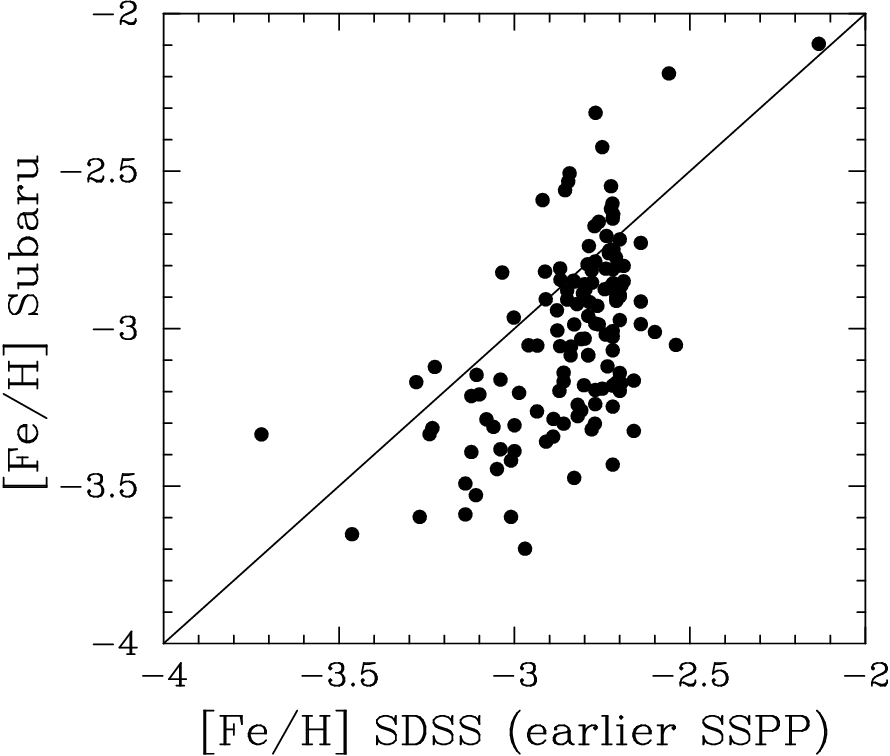}
\plotone{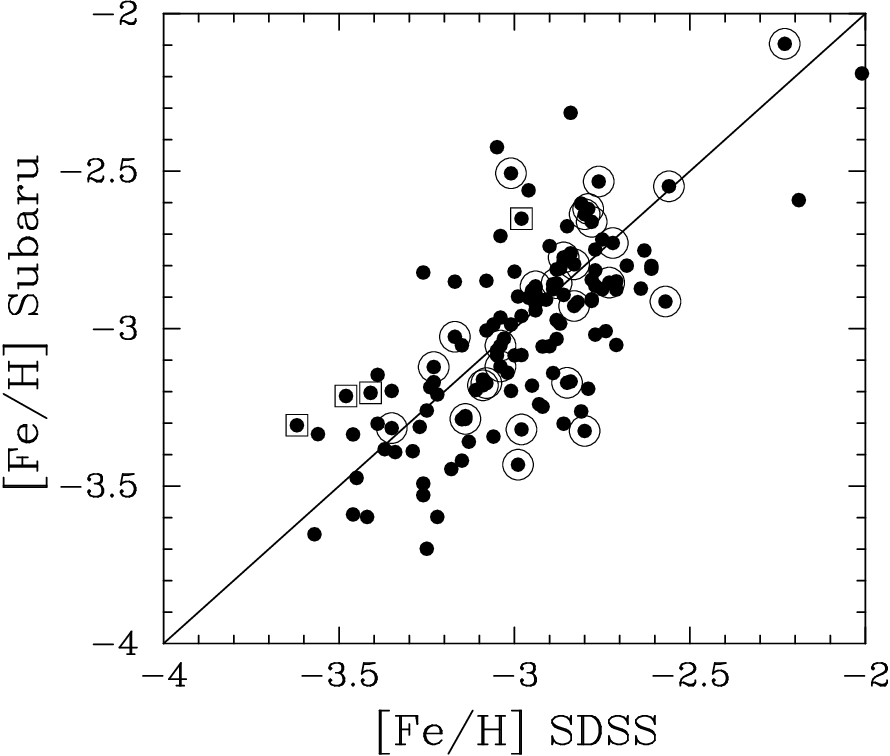}
\caption{{\it Upper panel:} Comparison of [Fe/H],
based on the snapshot Subaru high-resolution spectra, with those estimated
from medium-resolution SDSS spectra, based on the version of the SSPP in use prior 
to 2008. {\it Lower panel:} Same as the upper panel, but using
estimates from the latest version of the SSPP. Large open circles and
squares are over-plotted for giants and cool main-sequence stars,
respectively. The level of improvement in the SSPP is clear. \label{fig:compfe}}
\end{figure}

In order to assemble a set of likely EMP stars for high-resolution spectroscopy
with Subaru/HDS, we selected targets that have $V_{0} \lesssim 16.5$ ($g_0
\lesssim 16.7$) and [Fe/H]$\leq -2.7$, as provided by the SSPP, in the
temperature range 4500~K $<$ \teff\ $<$ 7000~K, over which the SSPP estimates
are best behaved. The choice of a conservative upper metallicity cut of
\feh\ $= -2.7$ was made because previous high-resolution follow-up
with Subaru and other telescopes had shown that the SSPP estimates of
metallicity, at the time of sample selection, were consistently 0.2-0.3
dex too high at the lowest metallicities. The upper panel of Figure
\ref{fig:compfe} shows the distribution of [Fe/H] estimated from SDSS
spectra (horizontal axis) using the version of the SSPP that was
available when the targets for Subaru/HDS observations were selected
in early 2008. Several stars having higher SSPP estimates of [Fe/H] or
$V_{0}>16.5$ were observed when appropriate targets did not exist in
the observing period with Subaru/HDS.  The Subaru high-resolution
estimates of \feh\ for these same stars, shown on the vertical axis,
are described below.

It is clear from Figure \ref{fig:compfe} that the conservative choice
on metallicity cut was indeed appropriate, as a considerable fraction
(65\%) of the stars with high-resolution estimates of \feh\ $< -3.0$
would have been missed had we set the selection boundary at
\feh\ (SSPP) = $-3.0$. The lower panel of Figure \ref{fig:compfe}
shows the effect of recent improvements in the SSPP, as discussed
further below (\S~\ref{sec:gmv}).

The list of 137 stars for which acceptable high-resolution spectroscopy was
obtained with Subaru/HDS is given in Table~\ref{tab:objlist}, where the object
name, coordinates, photometry, and reddening data are provided, and discussed in
detail below. In the following analysis, the objects are separated into turn-off
stars, giants, and cool main-sequence stars, based on determinations of their
effective temperature and gravity. According to this taxonomy, about 80\% of the
objects in our sample are main-sequence turn-off stars. Note that although
some objects were identified as carbon-rich stars from the SDSS
medium-resolution spectra prior to our obtaining high-resolution follow-up
spectra, we gave no preference in their choice (for or against), so that our
estimates of the fractions of such stars at low metallicity remains meaningful
(see below).

\section{Observations and Data Reduction}\label{sec:obs}

Acceptable quality high-resolution snapshot spectra for 137 of the
original 143 target stars selected above were obtained with the Subaru
Telescope High Dispersion Spectrograph \citep[HDS: ][]{noguchi02} in
four observing runs in 2008 (March, May, July, and October). Several
stars among the remaining eight stars were excluded because their
$S/N$ ratios were insufficient for our purpose. The other stars
exhibit spectra that differ from ``normal'' metal-poor stars. The
objects which are excluded in the analyses are reported in Appendix.

The spectra cover 4000--6800~{\AA}, with a gap of 5330--5430~{\AA} due to the
separation of the two EEV-CCDs used in the spectrograph. The resolving power
$R=36,000$ is obtained with the slit of 1.0 arcsec width and $2\times 2$ CCD
on-chip binning. The observing log is listed in Table~\ref{tab:obslog}, where
the observing dates, exposure time, S/N ratios, and heliocentric radial velocity
are presented. The average S/N ratios at 4300 and 5000~{\AA} per
resolution element are 31 and 51, respectively.

Data reduction was carried out with standard procedures using the IRAF
echelle package\footnote{IRAF is distributed by the National Optical Astronomy
Observatories, which is operated by the Association of Universities for Research
in Astronomy, Inc. under cooperative agreement with the National Science
Foundation.}, including bias-level correction using the over-scan regions of the
CCD data, scattered light subtraction, flat-fielding, extraction of spectra, and
wavelength calibration using Th arc lines. Cosmic-ray hits were removed by the
method described in \citet{aoki05}. Sky background was not very significant
in our spectra, which were obtained during times of little contamination from
the moon. The multi-order echelle spectra were combined to a single spectrum by
adding photon counts for the overlapping wavelength regions, and the combined
spectrum was then normalized to the continuum level. Spectra obtained with more
than one exposure were combined by adding photon counts before continuum
normalization.

\subsection{Equivalent Width Measurements}\label{sec:ew}

Equivalent widths ($W$'s) were measured for isolated absorption lines
in our spectra by fitting Gaussian profiles, using the line list given
in Table~\ref{tab:lines}. The measurements were made with a fortran
program of Gaussian fitting based on \citet{press92}, including an
estimate of the continuum level around each absorption line.

The number of lines detected in the spectra depends on the
metallicity, the stellar luminosity classification, and the $S/N$
ratio. In the spectra of turn-off stars, typically 10-20 \ion{Fe}{1}
lines are measured. In some spectra with relatively lower $S/N$ ratios
and relatively high temperatures, the number of \ion{Fe}{1} lines
detected is less than 10. While a few \ion{Fe}{2} lines are measured
in most turn-off stars, no \ion{Fe}{2} line is detected for 24 of our
stars. A few lines of \ion{Na}{1}, \ion{Mg}{1}, and \ion{Ca}{1} are
measured for most turn-off stars, while other elements (e.g., Sc, Ti,
Sr, Ba) are detected only for a limited number of objects. In general,
the number of lines measured for giants is larger than that for
turn-off stars, due to their lower effective temperatures. At least a
few \ion{Fe}{2} lines are detected for all giants in our sample.

\subsection{Radial velocities}\label{sec:rv}

\begin{figure}
\epsscale{.5}
\plotone{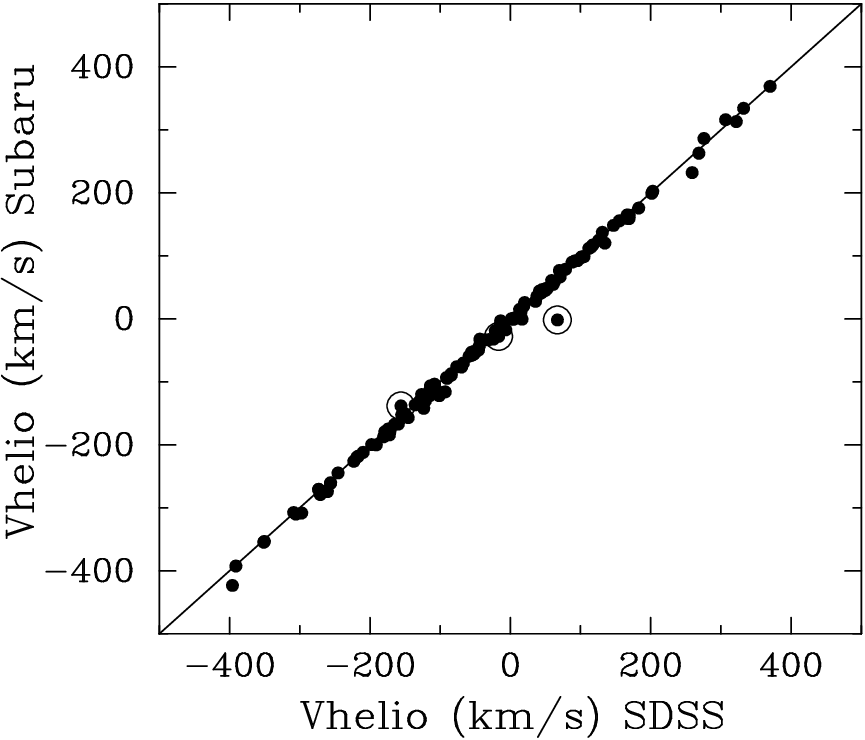}
\caption{Comparison of heliocentric radial velocities measured from
  the Subaru spectra with those from SDSS spectra. The open circles
  are over-plotted for the three double-lined spectroscopic
  binaries. Except for these three stars, the two measurements exhibit excellent
  agreement.
\label{fig:rv}
}
\end{figure}

Radial velocities are measured using the \ion{Fe}{1} lines for which
equivalent widths are measured. The derived heliocentric radial
velocities are given in Table~\ref{tab:obslog}. The random error in
the measurement is estimated to be $\sigma_{v} N^{-1/2}$, where $\sigma_{v}$
is the standard deviation of the derived values from individual lines,
and $N$ is the number of lines used. The table also provides the
values obtained from the SDSS spectra used for sample
selection. Comparisons of heliocentric radial velocities measured from
the Subaru spectra with those from SDSS are shown in
Figure~\ref{fig:rv}. In our sample, three double-lined spectroscopic
binaries are included, as reported below. The data points of these
stars are over-plotted by open circles in the figure. Excluding these
stars, the agreement between the two measurements is quite good, in almost all
cases well within the expected errors.

\section{Abundance Analysis}\label{sec:ana}

\subsection{Effective Temperature Estimates}\label{sec:param}

Chemical abundances are determined by a standard analysis for measured
equivalent widths using the ATLAS NEWODF grid of model atmospheres,
assuming no convective overshooting \citep{castelli03}. The
calculations of equivalent widths from models are made by the LTE
spectrum synthesis code based on the programs for the model atmospheres
developed by \citet{tsuji78}.

Owing to the lack of sufficient numbers of well-measured metallic
lines in our snapshot spectra, we are not in a position to determine
spectroscopic estimates of \teff\ by the usual practice of minimizing
the trend of the relationship between derived abundance and excitation
potentials of the lines from which it is derived. Balmer line
  profiles are also not used to determine {\teff}, because the S/N
  ratios of our data are too low for accurate estimation 
  of the continuum levels, although the Balmer lines were used in the
  first inspection of stellar types (see also Appendix).  Instead, we
have first estimated effective temperatures based on two sets of color
indicies, an approach that also has limitations. For example,
estimates of \teff\ derived for stars affected by large reddening are
more uncertain, due to errors in obtaining estimates of their
intrinsic colors.

Estimates of effective temperature based on $(V-K)_{0}$ colors are made 
using the temperature scales of \citet{casagrande10} for turn-off stars
({\teff}$_{\rm SSPP}\geq 5500$~K), and \citet{alonso99} for giants ({\teff}$_{\rm
SSPP}<5500$~K). The metallicity is assumed to be [Fe/H]$=-3.0$ for all stars in
order to carry out this calculation. The $V_{0}$ magnitude is derived from the
SDSS $g_{0}$ and $(g-r)_{0}$, using the transformations of
\citet{zhao06}, which are suitable for low-metallicity stars.
The $K_{0}$ magnitude is adopted from the 2MASS catalogue \citep{skrutskie06}.
In all cases, the absorption and reddening corrections were carried out
based on the reddening estimates from \citet{schlegel98}.

\begin{figure}
\epsscale{.42}
\plotone{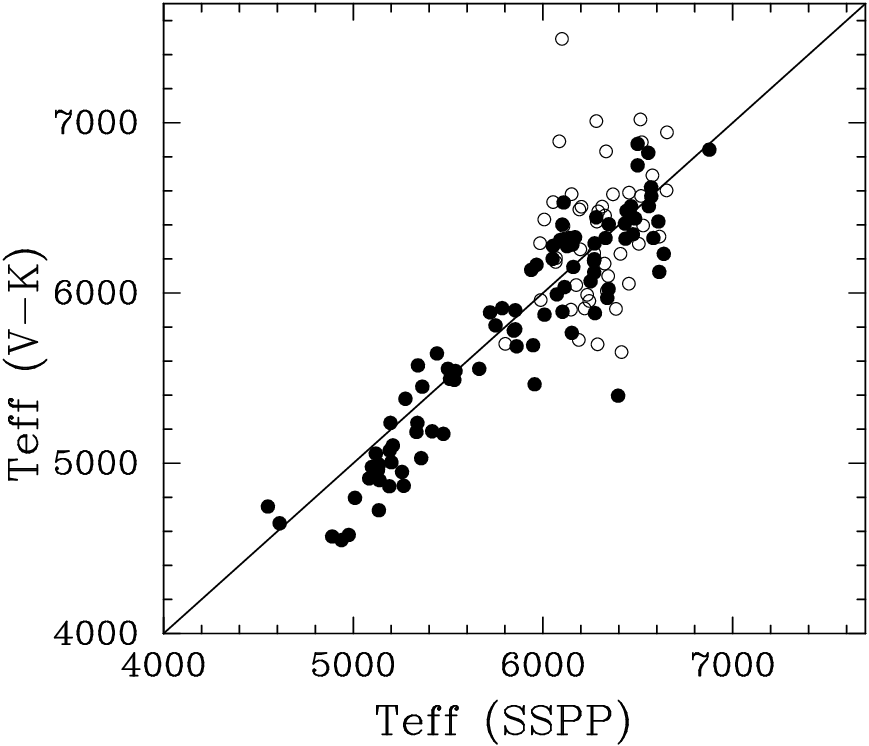} \\
\plotone{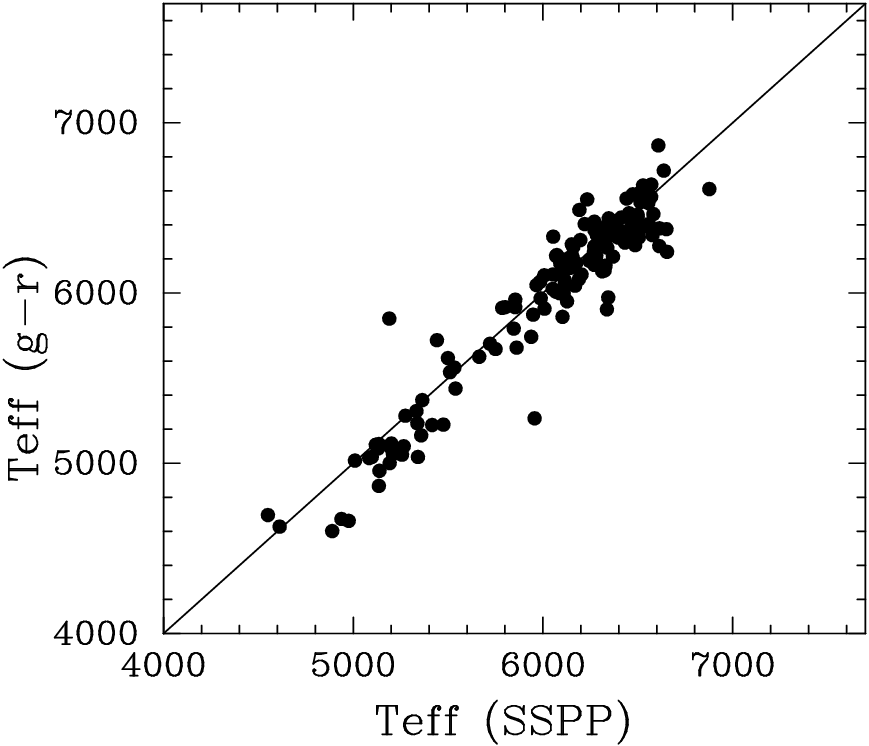} \\
\plotone{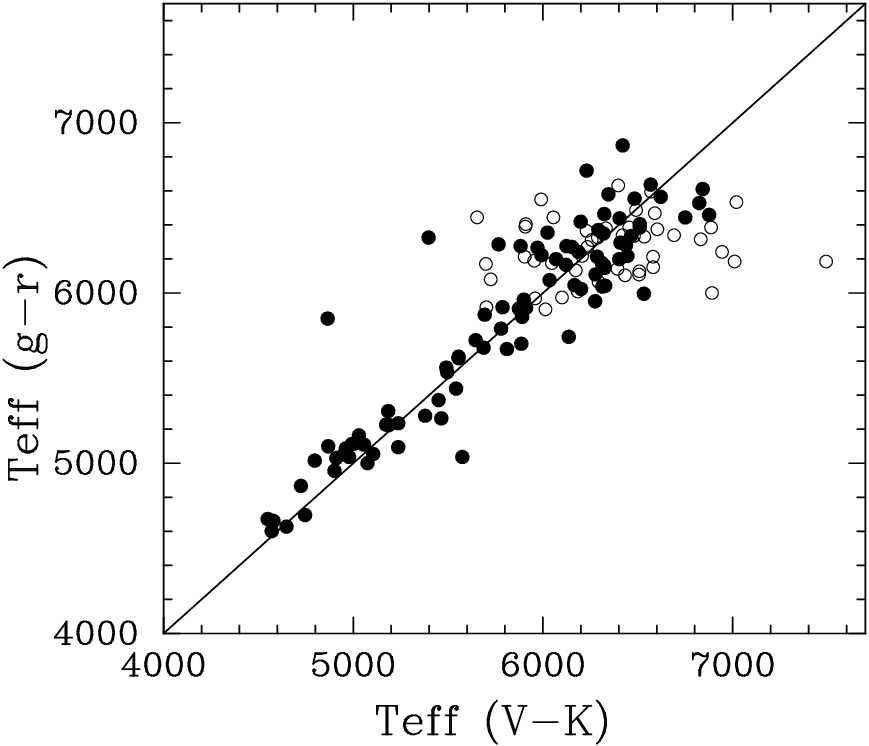} 
\caption{Comparison of {\teff} estimated by the (recent) SSPP and that
  from $(V-K)_0$ (top panel) and from $(g-r)_0$ (middle panel). The
  bottom panel shows the comparison of {\teff} between the two color
  indices. The open circles indicate objects with large errors in
  their $K$-band photometry. \label{fig:compteff}}
\end{figure}

The top panel of Figure~\ref{fig:compteff} shows a comparison of {\teff} between
the estimate from $(V-K)_{0}$ and that supplied by the SSPP. The SSPP actually
provides two sets of effective temperature estimates, one of which is based on
spectroscopy alone (essentially relying on the shape of the calibrated spectral energy
distribution, index measurements of temperature-sensitive lines, and spectral
fitting), while the other additionally includes photometric information in the
estimates. For the stars in our sample, we found essentially no zero-point
offsets between these approaches, with an rms variation of no more than 50~K.
Hence, we adopt the spectroscopy-only values determined by the SSPP for our
comparisons. Stars for which the {\teff} estimates from photometry are
potentially very uncertain, due to large reddening corrections, are excluded from
this comparison.

By inspection of the top panel from Figure~\ref{fig:compteff}, there is no significant offset between the two
estimates for turn-off stars ({\teff}$\gtrsim 5500$~K), although the scatter is
rather large. A likely cause of this scatter is the error in the $K$ apparent
magnitude measured by 2MASS, which can become large for stars as faint as some in
our sample. An error of 0.1~magnitude in $(V-K)_{0}$ results in a {\teff} error
of about 200~K. The 1$\sigma$ errors on the $K$ magnitude for some of our
fainter turn-off stars ($K>14.5$) are even larger than 0.1~magnitude. To draw
attention to these, they are shown as open circles in Figure~\ref{fig:compteff}.
The scatter in the {\teff} comparison for these stars is 395~K, much larger than
that for other stars of similar \teff\ having errors smaller than 0.1 in their
$K$ magnitudes ($\sigma=$241~K).

For the cooler stars ({\teff}$\lesssim 5500$~K), a clear offset (about 140~K) is
found between the {\teff} obtained from the $(V-K)_{0}$-based estimate and that
from the SSPP, although the rms scatter is small ($\sigma=$214~K). None of the
cooler stars have $K$ magnitude errors larger than 0.1 magnitude.

Estimates of {\teff} based on $(g-r)_{0}$ color are made using the
colors calculated based on ATLAS model atmosphere provided by Castelli
et
al.\footnote{http://wwwuser.oat.ts.astro.it/castelli/colors/sloan.html}. For
the purpose of this calculation, values of $\log g$ are assumed to be
4.0 and 2.0 for turn-off stars and giants, respectively, and the
metallicity is assumed to be [Fe/H]$=-3.0$. A comparison of these
{\teff} estimates with those supplied by the SSPP is shown in the
middle panel of Figure~\ref{fig:compteff}. There is no significant
offset between the two estimates for turn-off stars, and the scatter
is smaller (35~K) than that found for $(V-K)_{0}$ considered
above. Although a zero-point offset is found for the giants, it is
smaller (74~K) than that found for $(V-K)_{0}$. The reduced scatter
may be a result of smaller reddening effects on the $g-r$ color than
on $V-K$, or simply the better photometric precision of the measured
$g-r$ colors.

For completeness, the bottom panel of Figure~\ref{fig:compteff} compares the
effective temperatures estimated based on the two color indicies.

\begin{figure}
\epsscale{.6}
\plotone{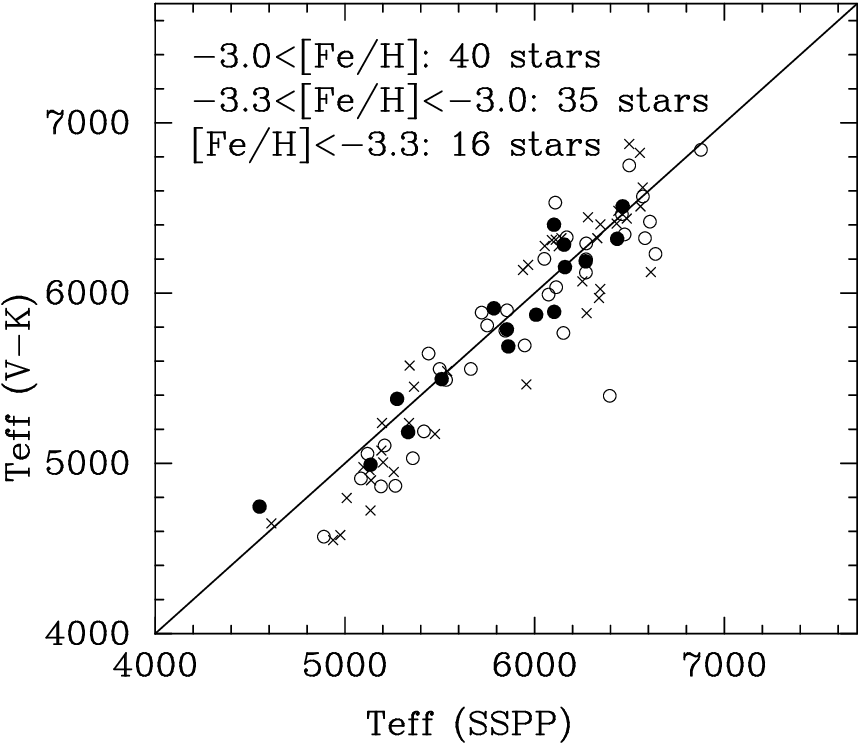}
\caption{Comparison of \teff\ , estimated by the (recent) SSPP, and
  those from $(V-K)_0$. The sample is separated into three metallicity
  ranges ([Fe/H] $<-3.3$: filled circles, $-3.3< $ [Fe/H] $ <-3.0$:
  open circles, and $-3.0 <$ [Fe/H]: crosses).\label{fig:compteffvk}}
\end{figure}

In order to assess the dependence of our {\teff} estimates on metallicity,
Figure~\ref{fig:compteffvk} provides comparisons for three metallicity ranges.
Note that we have separated this sample using the [Fe/H] values derived from the
abundance analysis in the present work, as described below. No significant
dependence on metallicity is seen.

Given the uncertainties in the $(V-K)_{0}$ based estimates, and the similarity of
the $(g-r)_0$ based estimates with those from the SSPP, we simply
adopt the (spectroscopic) effective temperatures determined by the SSPP
for the remainder of our analysis.

\subsection{Gravity, metallicity, and micro-turbulence}\label{sec:gmv}

The $\log g$ values for turn-off stars are expected to cover the range
from 3.5 (subgiant stars) to 4.5 (main-sequence stars), according to various
isochrones for very metal-poor stars
\citep{kim02,demarque04}. Unfortunately, for these stars, our
measurements only yielded a single or a few \ion{Fe}{2} lines, which
provides little opportunity to determine \logg\ in the traditional
manner, by demanding that the same abundance be returned by analysis
of the two ionization stages.  Hence, the surface gravity for turn-off
stars ({\teff}$>5500$~K) is simply set to $\log g=4.0$, and we accept
that errors in its determination can be as large as 0.5~dex.  We
assess the impact of this assumption below.

\begin{figure}
\epsscale{.70}
\plotone{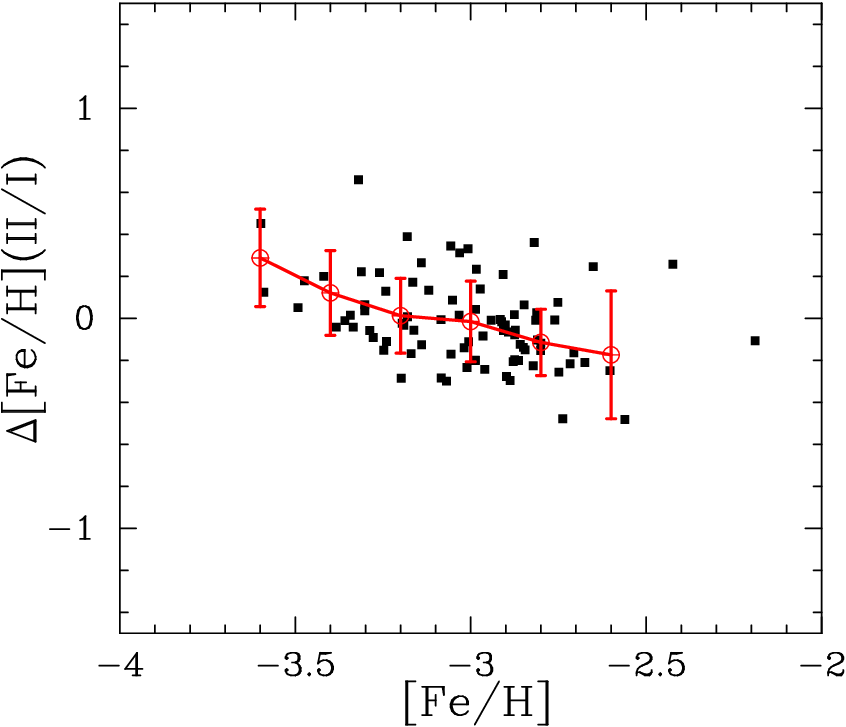}
\caption{The difference of [Fe/H] derived from \ion{Fe}{1} and
\ion{Fe}{2} for main-sequence turn-off stars (filled squares). The
average and standard deviation for each 0.2~dex bin of [Fe/H] are
represented by open circles and bars, respectively. The standard
deviations are about 0.20~dex, which is as small as the random errors
in the abundance measurements.  The lowest metallicity range likely
suffers from a bias, in that no \ion{Fe}{2} line is detected for a
larger fraction of stars than the less metal-poor stars. (See text for
details.)
\label{fig:dfe12}}
\end{figure}

Figure~\ref{fig:dfe12} shows the differences in [Fe/H] derived from \ion{Fe}{1}
and \ion{Fe}{2} for a sample of 88 turn-off stars for which at least one
\ion{Fe}{2} line is detected. The average and standard deviation for each
0.2~dex bin of [Fe/H] are represented by an open circle and bars,
respectively. No significant offset between the [Fe/H] abundances from the two
species appears for [Fe/H]$>-3.3$. The small offset found in the lower
metallicity range could be a result of a bias in the sample, arising from the
fact that no \ion{Fe}{2} line is detected for a larger fraction of these stars.
Indeed, \ion{Fe}{2} lines are detected for only 13 objects among the 24 turn-off
stars with [Fe/H]$<-3.3$. The \ion{Fe}{2} lines are weaker due to the generally 
higher gravities in objects for which the \ion{Fe}{2} lines are 
not detected. Excluding this bias, the gravity adopted in our analysis for
turn-off stars ($\log g =4.0$) is well justified, based on this comparison. 

The uncertainty in our $\log g$ values is estimated to be 0.5~dex, based on the
standard deviation of about 0.2~dex found in $\Delta$[Fe/H] from \ion{Fe}{1} and
\ion{Fe}{2} (see \S~\ref{sec:abana} and Table~\ref{tab:error}). We note that the iron
abundance measured from \ion{Fe}{1}, which is the metallicity indicator used in
this paper, as well as the abundance ratios [Mg/Fe], [Ca/Fe] and [Ba/Fe], which
are important for the discussion to follow, are not very sensitive to the $\log
g$ values.

We note that the {\logg} values determined by SSPP for our
  turn-off stars are lower than 4.0 on average, and those for 47 stars
  are lower than 3.5. We suspect that the uncertainty in gravity
  determination by the SSPP for EMP stars is still larger than the errors
  estimated for the entire SSPP sample (see \S~\ref{sec:sample1}), and further
  calibration for the lowest metallicity range is required.

For giants, where greater numbers of \ion{Fe}{2} lines are detectable, the $\log
g$ values are determined by seeking agreement between the iron abundances
derived from the \ion{Fe}{1} and \ion{Fe}{2} lines, within measurement errors.
Note that our sample includes four cool stars ({\teff}$\leq 5200$~K) that
exhibit very weak features of ionized species, including \ion{Fe}{2}, compared
to red giants of similar temperatures, indicating that they are on the
main sequence. Two of them (SDSS~1703+2836 and SDSS~2357--0052) have already
been studied by \citet{aoki10}. The surface gravities of these stars are
estimated by reading off the value from an isochrone appropriate for low-mass
metal-poor stars with ages of 12~Gyr \citep{kim02}, as was also done by
\citet{aoki10}. We adopt $\log g=4.8$ for the two stars studied by
\citet{aoki10}, and 5.0 for the other two cooler objects. The high surface
gravity for these stars explains the strong features of the \ion{Mg}{1} b lines,
due to the broader wings, as well as the detectable CH G-band, without assuming
exceptional over-abundances of C and Mg. 

The micro-turbulent velocity, {\vt}, for turn-off stars is fixed to 1.5~{\kms},
which is a typical value found by previous studies \citep[e.g., ][]{cohen04}.
Since the abundance measurements for most elements in turn-off stars are based
on weak lines, the result is insensitive to the adopted micro-turbulent
velocity. The values for giants are determined from the analysis of \ion{Fe}{1}
lines, by forcing the iron abundances from individual lines to exhibit no
dependence on the measured equivalent widths. The {\vt} of cool main-sequence
stars is assumed to be zero, which best explains the relation between the
equivalent widths of \ion{Fe}{1} lines and the abundances from individual lines.
There remains a weak trend in the relation, which can be resolved by assuming
negative values for {\vt}. This suggests that the line broadening from the
Uns\"{o}ld approximation should not be enhanced, as discussed by \citet{aoki10}.

The metallicity of the model atmospheres ([M/H]=[Fe/H]) is fixed to $-3.0$. The
temperature/density structure of photospheres, and the chemical abundances
derived using model atmospheres, is not very sensitive to the assumed
metallicity in such very/extremely metal-poor stars. Exceptions are found for
the three stars for which [Fe/H]$>-2.5$ was derived; in such cases we
iterated the analysis to obtain consistent [Fe/H] values.

Figure~\ref{fig:compfe} shows the [Fe/H] abundances determined by the
analysis of high-resolution spectra compared with those supplied by
the SSPP for SDSS medium-resolution spectra. The upper panel shows the
comparison for the estimates from SDSS spectra which were available
when target selection for the Subaru observations was carried out in
2008. This comparison indicates that very/extremely metal-poor stars
are efficiently selected from the SDSS measurements. Among the targets
for which [Fe/H] is estimated to be lower than $-2.7$ from the SSPP
estimate (most of objects in our sample, with few exceptions), only 10
stars have high-resolution determinations [Fe/H]$>-2.7$. This
demonstrates a clear advantage of using the results of the SDSS survey
for picking EMP targets, relative to previous surveys based on
low-resolution objective-prism spectra.  

On the other hand, the correlation between the two measurements of
[Fe/H] is weak, as seen in the upper panel of
Figure~\ref{fig:compfe}. In particular, stars for which the iron
abundance is estimated to be $-3.0<$[Fe/H]$<-2.7$ exhibit very large
scatter. Since the number of objects that have high [Fe/H] ([Fe/H]$>-2.5$;
measured from high-resolution spectra) is small, the large scatter is
due to the existence of many EMP ([Fe/H]$<-3.0$) stars among
them. Moreover, almost all stars for which [Fe/H]$<-3$ as derived from
the SDSS spectra exhibit lower [Fe/H] as obtained from the
high-resolution spectra. This comparison indicates that the criterion
in the sample selection for the [Fe/H] values from SDSS (based on an
earlier version of SSPP) provides a homogeneous sample for lower
metallicity ([Fe/H]$\lesssim -3$), while the selection could be
incomplete for stars of higher metallicity.

The lower panel of Figure~\ref{fig:compfe} shows a comparison of
[Fe/H] abundances derived by our analysis of high-resolution spectra
with the SDSS estimates supplied by the latest version of SSPP. In
contrast to the upper panel, there is no clear offset between the two
estimates. However, the scatter is still larger than preferred for
detailed inference of, e.g., the nature of the metallicity
distribution function (MDF) at very/extremely low metallicity. We also
note that, in particular for metallicities as low as those considered
in our present analysis, the effects of interstellar \ion{Ca}{2} on
metallicities calculated from medium-resolution spectra can be large,
in particular for warmer stars, which rely almost entirely on the
strength of the \ion{Ca}{2} K line for their metallicity
estimation. Thus, we are reminded once again that high-resolution
spectroscopy is required to obtain accurate metallicity estimates for
individual EMP stars.

The [Fe/H] values we estimate for the four cool main-sequence stars from our
high-resolution spectra are lower than those from the SSPP, by $\sim 0.3$~dex.
The gravity estimates for these stars are also significantly higher than
reported by the SSPP for the medium-resolution spectra, which may also
contribute some to the offset in [Fe/H]. It should be kept in mind that, due to
their low luminosities, cool dwarfs infrequently enter into samples of VMP/EMP
stars selected from magnitude limited samples. Nevertheless, future adjustments
to the SSPP may be able to better handle such stars.

\subsection{Abundance measurements}\label{sec:abana}

Standard LTE abundance analyses for measured equivalent widths have
been made for other elements. The derived abundances are presented in
Table~\ref{tab:abundance}. The errors given in the table include
random errors and those due to uncertainties of atmospheric
parameters. The random errors in the measurements are estimated to be
$\sigma N^{-1/2}$, where $\sigma$ is the standard deviation of derived
abundances from individual lines, and $N$ is the number of lines
used. Since the $N$ for most elements other than Fe is small, 
the
$\sigma$ of \ion{Fe}{1} ($\sigma_{\rm Fe I}$) is adopted in the
estimates for them (i.e. error is $\sigma_{\rm Fe I} N^{-1/2}$). The
errors due to the uncertainty of the atmospheric parameters are
estimated for a turn-off star and a giant (Table~\ref{tab:error}), for
$\delta${\teff}$= 150$~K, $\delta${\logg}$=0.5$, and $\delta${\vt}$=
0.5$~{\kms}). These errors are added in quadrature to the random
errors to derive the total errors given in Table~\ref{tab:abundance}.

\subsection{Carbon abundances}\label{sec:ana_c}

\begin{figure}
\epsscale{.50}
\plotone{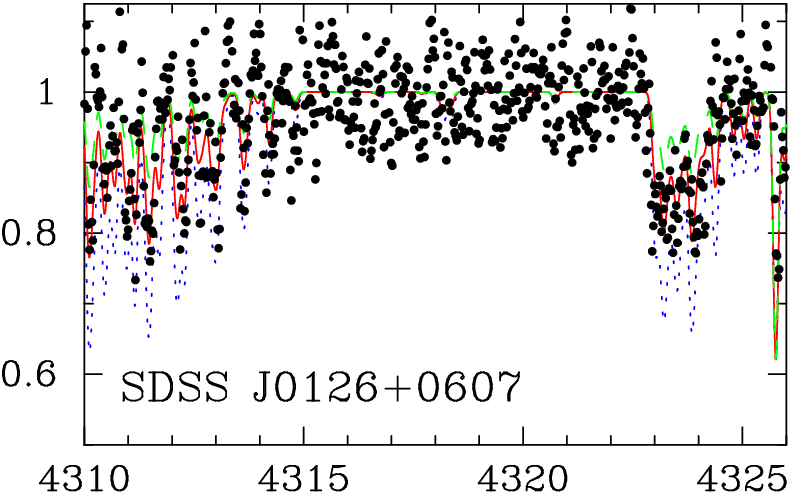}
\plotone{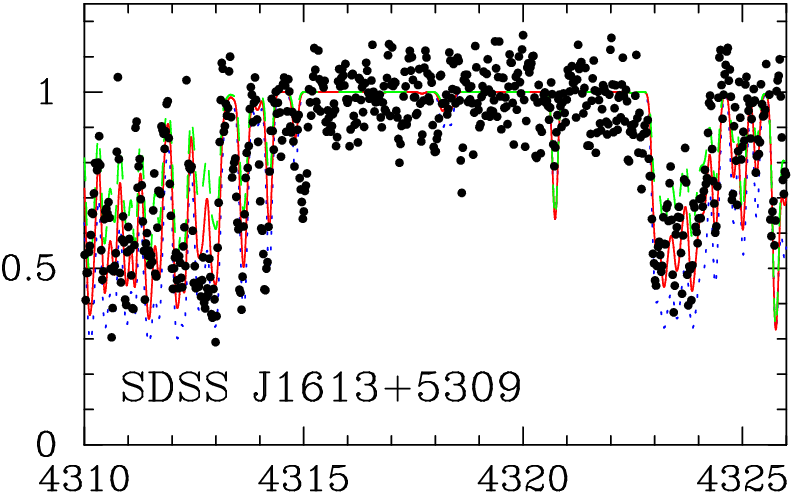}
\plotone{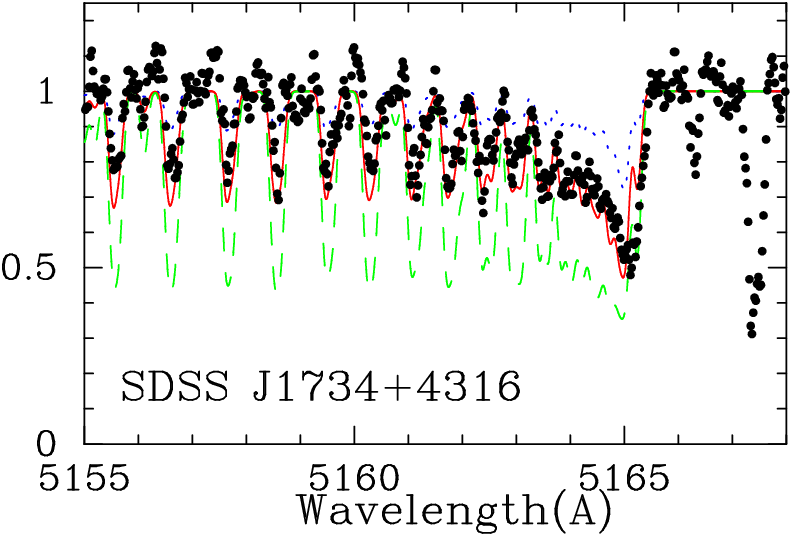}
\caption{Spectra of the CH G-band/4323~{\AA} band and the C$_{2}$ Swan
  band at 5165~{\AA}, for the stars labeled in each panel (filled
  circles). The three synthetic spectra are calculated by changing
  [C/Fe] by 0.3~dex (dashed and dotted lines) around the adopted
  values (solid line).
\label{fig:chc2}}
\end{figure}

Carbon abundances are determined for 28 stars in our sample, based on
the CH G-band (the Q branches of CH A--X system at $\sim 4320$~{\AA}), as
well as the C$_{2}$ Swan 0--0 band at 5165~{\AA}. Examples of the
spectra of the CH band are shown in Figure~\ref{fig:chc2}. The molecular
data are the same as those used in \citet{aoki07}.

The CH G-band is detected for 14 stars among the 25 giants, and for
all four of the cool main-sequence stars. Nine giants are
carbon enhanced ([C/Fe] $> +0.7$).  The C$_{2}$ band is used to
determine the carbon abundances of four stars, because the CH G-band
in these objects is almost saturated. The detection limit of the CH
G-band in a red giant with {\teff} $ < 5500~$K is approximately
[C/H]$\sim -2.5$ for spectra of snapshot quality (S/N$\sim
30$). Hence, stars for which the CH G-band feature is not detected are
unlikely to be CEMP stars. The fraction of CEMP stars in our sample is
9/25$\sim$ 36\%, which is in agreement with the estimate by Carollo et
al. (2012) for [Fe/H]$\sim -3$, i.e., at the limit of their sample.

None of the four cool main-sequence stars in our sample are carbon enhanced. The
CH G-band is detected for such objects because of their low temperatures and the
high pressure of their atmospheres. Among the four stars, the lowest carbon
abundance ratio is found for SDSS J0018-0939 ([C/Fe]$=-0.7$). Although the CH
G-band of this star is weak and the measurement is uncertain, a conservative
upper limit is [C/Fe]$<-0.3$, which is already lower than the carbon abundances
found in other cool main-sequence stars. A weak CH G-band could alternatively be
explained by assuming a lower gravity. However, the weak features of ionized
species such as \ion{Fe}{2} cannot be accounted for if this object is assumed to
be a red giant ($\log g<3$). Thus, the under-abundance of carbon in this star is
a robust result, although a higher quality spectrum is required to derive an
accurate estimate of its abundance.

In contrast to the giants, the CH G-band is detected for only 10 stars
among the 108 turn-off stars. All of these objects are highly
carbon enhanced ([C/Fe] $\gtrsim +2$). The detection limit of the CH
G-band for a turn-off star with {\teff}$\sim 6000$~K is [C/H]$\sim
-1.5$ ([C/Fe]$\sim +1.5$ for [Fe/H]$\sim -3$), indicating that the CH
G-band is not measurable even for the mildly carbon-enhanced ($+0.7 <
$[C/Fe]$ < +1.5$) stars in our sample. Hence, the fraction of CEMP
objects for the turn-off stars (10/108$\sim$ 9\%) should be regarded
as a lower limit.

\subsection{Analysis of the comparison star G~64--12}\label{sec:comp}

In order to examine the reliability of the abundances determined from
our snapshot spectra, we obtained a spectrum of the well-studied EMP
turn-off star G~64--12, using the same instrumental setup and
integrating to a similar S/N ratio as for the bulk of our sample,
employing a short (five minute) exposure time. The chemical
composition of this object was reported on by \citet{aoki06}, using a
high-resolution, high S/N spectrum. The abundance analysis for this
object was conducted adopting the same model atmosphere used in
\citet{aoki06}, that is, the ATLAS model including convective
overshooting \citep{kurucz93} for {\teff} $=6380$~K, $\log g=4.4$ and
[Fe/H] $ = -3.2$. The results are compared in
Table~\ref{tab:comparisonstar}. The [Fe/H] and [X/Fe] values are
calculated adopting the solar abundances of \citet{asplund09} for both
cases. The agreement is fairly good for most species: the differences
of the two measurements ($\log \epsilon$ values) are within 0.12~dex,
which are as small as the random errors in the present analyses.  An
exception is the Sr abundance, which is determined from the
\ion{Sr}{2} resonance lines in the blue range, where the data quality
is relatively low. Another exception is Na, for which only a
preliminary result was provided in \citet{aoki06}. The Na abundance
was determined from the same spectrum to be $\log \epsilon$(Na)
$=2.74$ by \citet{aoki09}, adopting slightly different atmospheric
parameters (Table~\ref{tab:comparisonstar}), which agrees well with
the Na abundance measured by the present work ($\log \epsilon$(Na)
$=2.82$).

\subsection{Double-lined spectroscopic binaries}\label{subsec:sb2}

\begin{figure}
\epsscale{.70}
\plotone{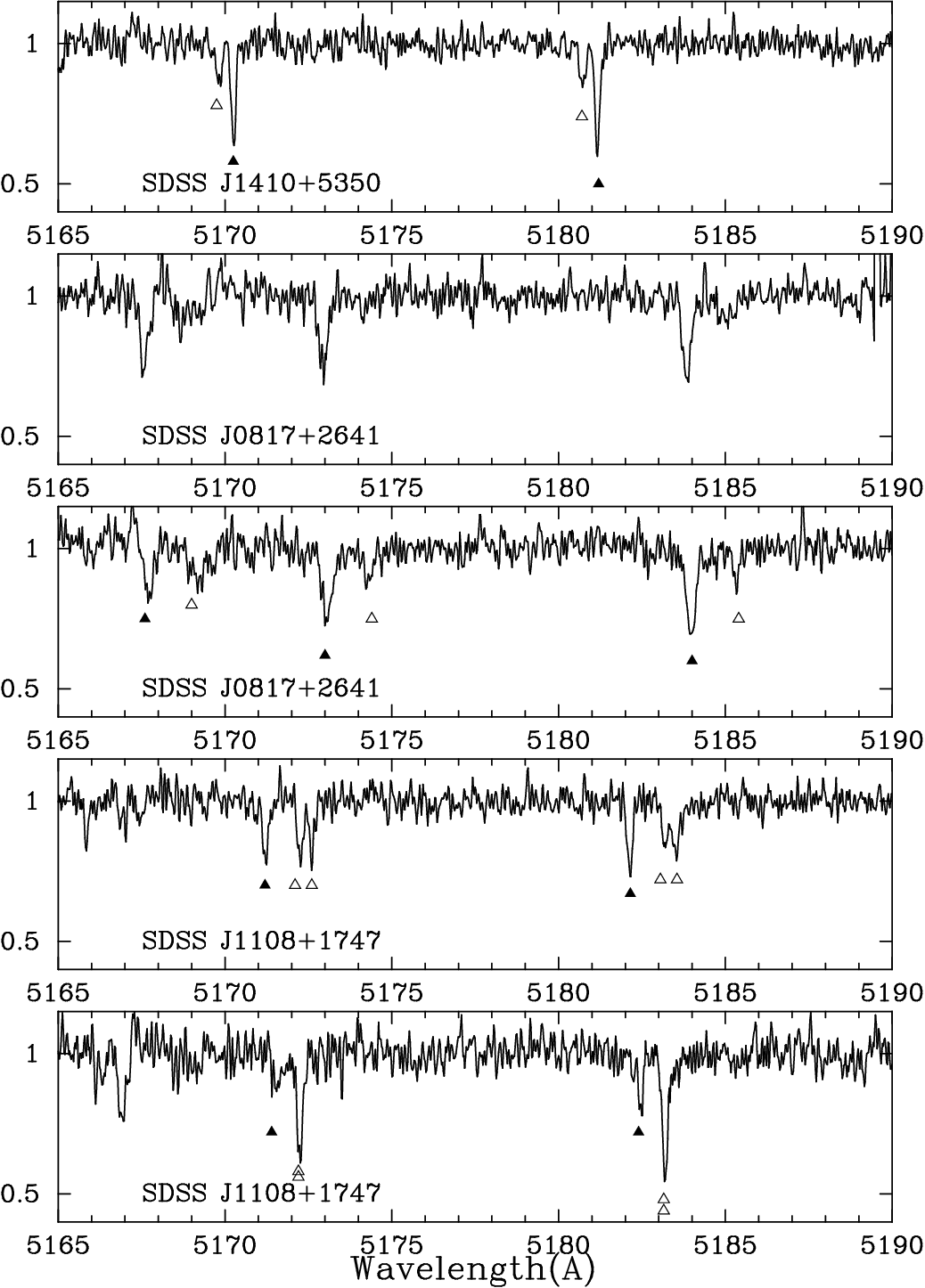}
\caption{Spectra around the Mg b lines of double-lined and
  triple-lined spectroscopic binaries: SDSS J1410+5350 (the top
  panel), SDSS J0817+2641 (second and third panels), and SDSS
  J1108+1747 (the lower two panels). The filled triangles indicate the
  primary \ion{Mg}{1} b lines for each star, while the open triangles
  show the secondary ones. \label{fig:sb2}}
\end{figure}

Three stars in our sample clearly exhibit two (or three, in the case of one
star) sets of absorption features with distinct doppler shifts, suggesting these
objects to be double-lined spectroscopic binaries or multiple systems. The
region of the spectra around the Mg b lines of these objects is shown in
Figure~\ref{fig:sb2}. 

SDSS J0817+2641 was already studied by \citet{aoki08}. The spectrum
obtained in their study (on February 9, 2007) is shown in the second
panel of Figure~\ref{fig:sb2}.  The second component of the absorption
features is not obvious in their spectrum, while it is clearly seen in
our new spectrum shown in the third panel.  \citet{aoki08} reported a
discrepancy in the radial velocities measured from the SDSS
medium-resolution spectrum and their Subaru high-resolution data. The
discrepancy is most likely a result of the large doppler shifts of
both components, or possibly due to the uncertainty of the
measurements from the medium-resolution spectrum, which was incapable
of resolving the two components.  This also suggests that, given the
overall excellent quality of the SDSS stellar velocity determinations
(especially for the brighter stars one naturally targets for
high-resolution spectroscopic follow-up), as described in
\S~\ref{sec:rv}, that obtaining even a single-epoch snapshot quality
high-resolution spectrum is an efficient way to identify at least
high-amplitude binary variations in a sample of stars.

The spectrum of SDSS J1108+1747, obtained on March 7, 2008 (fourth panel of Fig.
\ref{fig:sb2}) exhibits triple spectral features, indicating that this system
consists of at least three stars of similar luminosity. A significant change of
the spectral features is found in its March 9 spectrum (fifth panel), in which
only two components appear. The correspondence between the features of the two
spectra is still unclear because of the limited quality of our spectra. This is,
to our knowledge, the most metal-deficient multiple ($n \geq 3$) system yet
discovered ([Fe/H] $\sim -3$).

In order to accurately measure the chemical compositions of these
objects, we need to determine the contribution of each component to
the continuum light.  This is only possible by determining the mass
ratios from long-term radial velocity monitoring. However, the ratios
of the apparent strengths of the Mg b lines (depths of the absorption
lines) are at most three or four, suggesting that the components have
comparable luminosities. The objects have {\teff} around
6000~K. Hence, the spectral features can be modeled by adding the
spectra of two or three main-sequence stars that have slightly
different mass (mass ratios of 1.2 or smaller; see Goldberg et
al. 2002), which have similar strengths of the (partially saturated)
Mg b lines. For a rough estimate of the contribution of each component
to the continuum light, we assume that the contribution is in
proportion to the depth of the Mg b absorption lines. The primary
components of the Mg b lines identified by the present work are
indicated by the filled triangles in Figure~\ref{fig:sb2}. The
contributions of the primary components estimated by this method are
75\%, 80\%, and 40\% in SDSS J1410+5350, SDSS J0817+2641, and SDSS
J1108+1747, respectively. The equivalent widths measured for the
primary components by Gaussian fitting are divided by these factors
for carrying out the abundance analyses.

The abundance analyses for the primary components of these objects are made as
for other single-lined stars. We adopt the {\teff} determined by SSPP with no
modification; the {\teff}'s estimated from colors agree with the SSPP results.
The colors of the (integrated) system should be similar to the colors of the
primary component -- if the primary is distinctively warmer than the other
components, it should dominate the system luminosity, while the primary should
have similar {\teff} to the other components if it does not dominate the
system luminosity.

\section{Discussion}\label{sec:abratio}

\subsection{Carbon-Enhanced Metal-Poor (CEMP) Stars}\label{sec:cemp}

Previous studies of CEMP stars have clearly demonstrated that they are separable
into at least two primary classes: CEMP stars exhibiting large enhancements of
s-process elements (CEMP-s), and those with no excess of neutron-capture
elements (CEMP-no) \citep{beers05}. Note that recent studies of CEMP-s stars
split this class even more granularly (e.g., Bisterzo et al. 2011). The fraction
of CEMP-no stars among all CEMP stars is $\sim 20$\%, and increases with
decreasing stellar metallicity \citep{aoki07, hollek11}. The distribution of the
[C/H] ratios also appears to be different between the two classes -- most
of the CEMP-s stars exhibit quite high values ([C/H]$ > -1$), while the CEMP-no
class exhibits a wide distribution of values \citep{aoki07}.

Among the nine CEMP red giants in our sample, only two possess large excesses of
Ba ([Ba/Fe] $ > +1$: SDSS J1836+6317 and SDSS J1734+4316). One exhibits a
moderate excess of Ba ([Ba/Fe]$=+0.8$: SDSS J0711+6702). Four stars have solar
or lower Ba abundance ratios ([Ba/Fe]$\lesssim 0$), hence are classified as
CEMP-no stars. Although Ba is not measured for the other two CEMP red giants,
they could also be CEMP-no, given the detection limit of Ba in red giants
([Ba/Fe] $\sim -0.5$ at [Fe/H] $\sim -3$). Hence, six objects among the nine
CEMP giants in our sample are CEMP-no stars. This result suggests that a high
fraction of CEMP-no stars exists in this metallicity range ($-3.4 < $[Fe/H]$ <
-2.5$). 

Possible progenitors for the CEMP-no class include massive, rapidly
rotating, mega metal-poor ([Fe/H] $< -6.0$) stars, which models
suggest have greatly enhanced abundances of CNO due to distinctive
internal burning and mixing episodes, followed by strong mass loss
(Hirschi et al. 2006; Meynet et al. 2006, 2010a,b). Another suggested
mechanism for the production of the material incorporated into CEMP-no
stars is pollution of the interstellar medium by so-called faint
supernovae associated with the first generations of stars, which
experience extensive mixing and fallback during their explosions
(Umeda \& Nomoto 2003, 2005; Tominaga et al. 2007). This model
well-reproduces the observed abundance pattern of the CEMP-no star
BD+44$^{\circ}$493, the ninth-magnitude [Fe/H] $= -3.7$ star (with
[C/Fe] = +1.3, [N/Fe] = +0.3, [O/Fe] = +1.6) discussed by Ito et
al. (2009). The recently reported high redshift ($z = 2.3$), extremely
metal-poor Damped Lyman-$\alpha$ (DLA) system by Cooke et al.  (2011:
[Fe/H]$\sim -3.0$) exhibits enhanced carbonicity ([C/Fe] $= +1.5$) and
other elemental abundance signatures that \citet{kobayashi11} also
associate with production by faint supernovae. In addition, a fraction
of CEMP-no stars might belong to binary systems and have been formed by mass
transfer from AGB stars that yielded no s-process elements but enriched
carbon \citep[e.g., ][]{suda04}.

Eight of the 10 CEMP turn-off stars exhibit large enhancements of
Ba. Although the metallicities of these stars are similar to those of
the CEMP giants, the fraction of CEMP-s stars is apparently higher
among the turn-off stars than for the red giants. The carbon
over-abundances of all the CEMP turn-off stars are much larger than
the average over-abundances of the CEMP giants, presumably because the
detection limit of the CH G-band is higher for these warmer stars. As
shown by previous studies of CEMP stars \citep[e.g., ][]{aoki07}, the
[C/H] distributions are quite different between the two classes of
CEMP stars: a large fraction of CEMP-s stars have higher [C/H]
values. Hence, the high fraction of the CEMP-s stars among the CEMP
turn-off stars could be simply due to this bias in the sample. In
other words, one can assume that many additional CEMP-no stars could
be included among the turn-off stars of our sample, but they have not
yet been identified as CEMP stars.

We comment here on five CEMP stars that have remarkable features in their
chemical compositions or stellar parameters.

{\it SDSS J1036+1212 -- a CEMP-s star with [Fe/H]$=-3.5$ and a Spite
  plateau Li abundance}: This extremely metal-poor turn-off star
({\teff}$=5850$~K) exhibits excesses of carbon ([C/Fe] $=+1.5$) and Ba
([Ba/Fe] $=+1.3$). The [Sr/Ba] ratio is very low ([Sr/Ba] $=-2.1$),
suggesting the contribution of the s-process even at extremely low
metallicity, which efficiently produces heavy elements due to the high
ratio of neutrons to seed nuclei \citep[e.g., ][]{busso99}. The
\ion{Li}{1} resonance line is clearly detected for this star, even in
our snapshot spectrum with a moderate S/N ratio. The equivalent width
of the Li line is 52m{\AA}, resulting in $\log
\epsilon$(Li)$=2.2$. This value agrees with the Li abundances
typically found for metal-poor turn-off stars (the so-called Spite
plateau value: e.g., Spite \& Spite 1982; Mel{\'e}ndez et
al. 2010). The excesses of carbon and barium in such stars is usually
interpreted as the result of mass transfer from a companion AGB
star. If the amount of mass transferred from the AGB star was large,
and the Li was depleted on the surface of the AGB star, the Li should
also be depleted in the star we are currently observing. Thus, more
accurate determinations of the Li abundance of this star will provide
an upper limit on the mass transferred from the AGB star. We have
already obtained a higher S/N spectrum of this object, and a detailed
study will appear in a separate paper in this series.

{\it SDSS J1613+5309 -- a CEMP-no star with a Mg excess}: This object
is an EMP ([Fe/H] $=-3.3$) giant that exhibits a moderate excess of Mg
([Mg/Fe] $=+0.9$), but no excess of Ba. Two previously identified CEMP
stars are known to possess large excesses of $\alpha$-elements
(CS~22949--037 and CS~29498--043: McWilliam et al. 1995, Aoki et
al. 2002a, Aoki et al. 2002b), and are referred to as CEMP-$\alpha$
stars \citep{beers05}. Although the excess of the $\alpha$-elements of
SDSS J1613+5309 is not as clear as for the two previous objects, this
star is likely a new member of the CEMP-$\alpha$ class. We note in
passing that enhanced $\alpha$-elements are often (though not always)
associated with the CEMP-no class.

{\it SDSS J1836+6317 and SDSS J1245$-$0738 -- CEMP-s stars with large
  excesses of Na and Mg}: These are typical CEMP-s stars, having
[Ba/Fe] $>+2.0$. They exhibit large excesses of Na ([Na/Fe] $>+1.3$)
and moderate excesses of Mg ([Mg/Fe] $\sim +0.8$). Similar
over-abundances of Na and Mg are also found in several previously
studied CEMP-s stars, e.g., LP~625--44 \citep{aoki02a}.  Although the
source of the Na and Mg in such objects is not well understood,
nucleosynthesis in AGB stars that yielded the large over-abundances of
neutron-capture elements may also be related to the production of
these light elements. For instance, the s-process models by
\citet{bisterzo11} suggest dependence of Na production by
$^{22}$Ne($n,\gamma$)$^{23}$Ne (and $\beta$-decay of $^{23}$Ne) on
stellar mass.

{\it SDSS J0126+0607 -- a ``hot'' CEMP-s star}: The {\teff} of this
object is the highest (6900~K) in our sample, and the excesses of
carbon and Ba are also the highest ([C/Fe] $=+3.1$ and [Ba/Fe]
$=+3.2$). Such CEMP-s stars could be formed by accretion of
significant amounts of carbon-enhanced material from an AGB star
across a binary system. Some previously known hot ({\teff} $\sim
7000$~K) CEMP-s stars (e.g., CS~29497--030: Sivarani et al. 2004,
Ivans et al. 2005; CS~29526--110: Aoki et al. 2008) exhibit variations of
radial velocities with short timescales (less than one year). Future
monitoring of the radial velocity for this object, as well as more
detailed chemical-abundance studies, will provide new insight for the
formation mechanism of such hot CEMP-s stars.

\subsection{The $\alpha$-elements}\label{sec:alpha}

\subsubsection{Abundance scatter and outliers}

\begin{figure}
\epsscale{.42}
\plotone{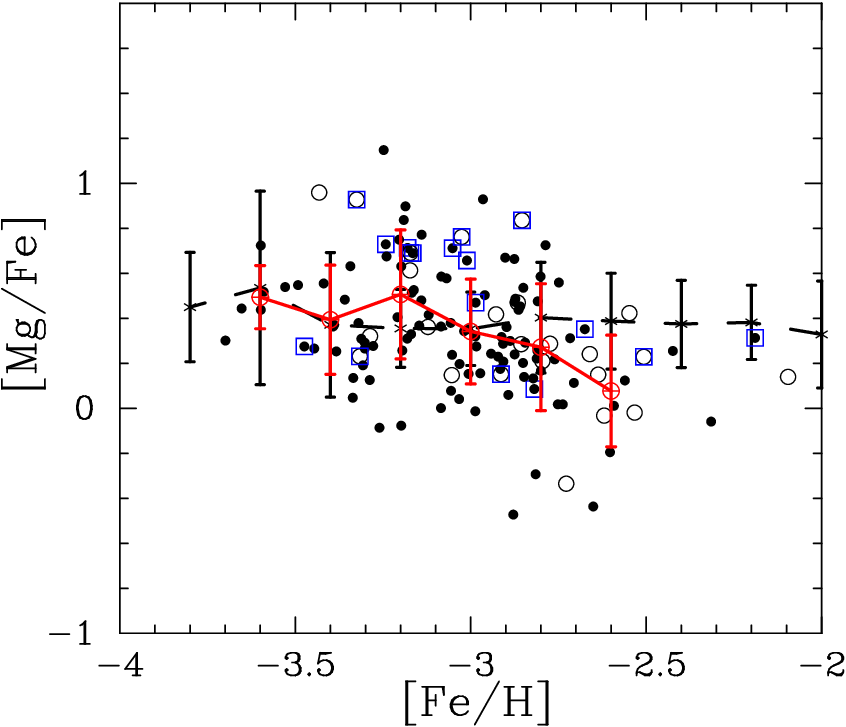} \\
\plotone{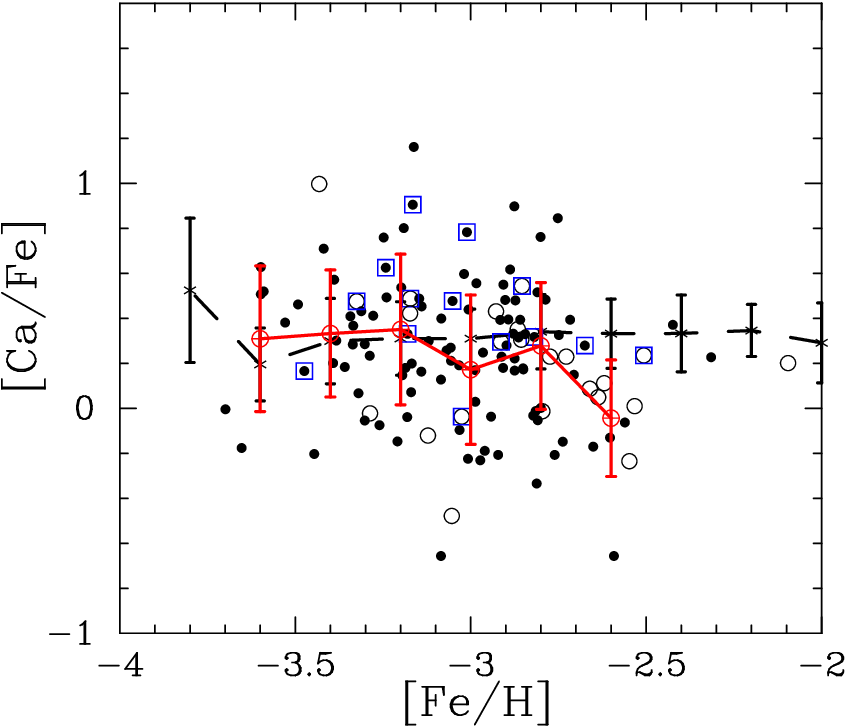} \\
\plotone{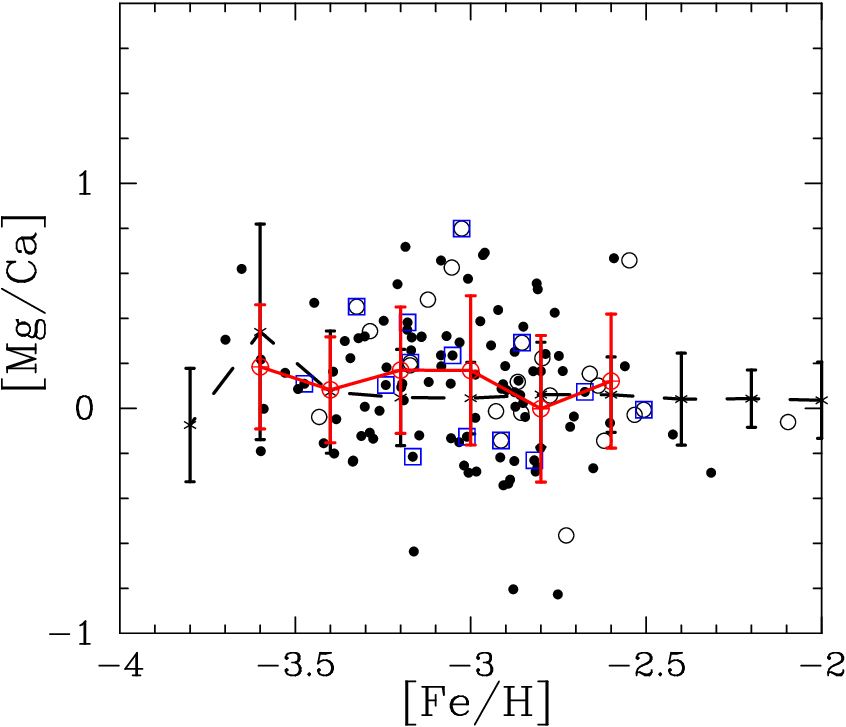} 
\caption{Abundance ratios of [Mg/Fe] (top panel), [Ca/Fe] (middle
  panel), and [Mg/Ca] (bottom panel), with respect to [Fe/H]. The
  filled and open circles indicate main-sequence turn-off and giant
  stars, respectively. CEMP stars ([C/Fe] $>+0.7$) are shown by
  over-plotting open squares. Large open circles and bars, connected
  by a solid line (red), represent the average and standard deviation
  of the abundance ratios, respectively, for each metallicity bin of
  width 0.2 dex for our SDSS sample. The standard deviations are
  0.20--0.25~dex, which are as small as the random errors in the
  abundance measurements. Crosses and bars, connected by a dashed
  line, indicate those for the SAGA sample (see text).
\label{fig:mgca}}
\end{figure}

Figure~\ref{fig:mgca} shows the abundance ratios of [Mg/Fe], [Ca/Fe],
and [Mg/Ca] for turn-off stars (filled circles) and cooler stars (open
circles). The average and standard deviation of the abundance ratios
for each 0.2~dex bin of [Fe/H] are indicated by large open circles and
bars, connected by a solid line. The average and standard deviation of
the abundance ratios determined by previous studies, which are taken
from the SAGA database \citep{suda08}, are also shown by crosses and a
dashed line, for comparison. The standard deviations of
[Mg/Fe] and [Ca/Fe] of our sample are 0.25--0.35~dex, as small as the
measurement errors. Hence, no clear intrinsic scatter of the abundance
ratios is found in these diagrams.

Non-LTE effects on abundance measurements for extremely
  metal-poor stars were investigated for Mg by \citet{andrievsky10},
  and for Ca by \citet{mashonkina07} and \citet{spite12}. The non-LTE
  corrections for Mg are positive, and the size is 0.1--0.3~dex,
  according to \citet{andrievsky10}. The corrections are
  systematically larger for dwarf stars than for giants. Hence, the Mg
  abundance ratios could be systematically higher than those derived
  by our LTE analysis. The non-LTE corrections for Ca abundances
  measured from neutral species are dependent on spectral lines: the
  correction is largest for measurements based on the \ion{Ca}{1} 4226~{\AA}
  line, which is used in our analysis for a portion of the sample. The
  corrections are estimated to be 0.0--0.2~dex by
  \citet{spite12}. Hence, the effects are not significant when other
  lines (such as the subordinate lines) of \ion{Ca}{1} are available, as in the
  case of analyses for giants in our sample. We note that the
  abundance results taken from the SAGA database are based on LTE
  analyses.

We would like to comment on possible outliers in Mg abundances as candidate
$\alpha$-element-enhanced or $\alpha$-deficient stars. Although some apparent outliers
exist in [Ca/Fe], the Ca abundances are determined from only one feature at
4226~{\AA} in many cases, for which the S/N ratios are not as high as in the red range, and
which could be affected by contamination from the CH features in carbon-enhanced
objects. Hence, the Ca abundances are discussed only for comparison purposes.

There are seven stars that have [Mg/Fe] $>+0.8$. Two of them are CEMP stars, as
discussed above. Three objects (SDSS J0840+5405, SDSS J1623+3913, and SDSS
J2104$-$0104) also exhibit some excess of Ca ([Ca/Fe] $>+0.7$), suggesting that
their excesses of the $\alpha$-elements are real.

The Mg abundance ratios of the other two stars (SDSS J1412+5609 and
SDSS J1424+5615) are also very high ([Mg/Fe]$\sim +0.9$), while their
Ca abundances appear normal. If this result is real, this suggests
scatter of chemical-abundance ratios produced by the progenitor
massive star and its supernova explosion.

There are four stars that have [Mg/Fe] $<-0.2$. One of the four stars is a cool
main-sequence star (SDSS J0018$-$0939: [Mg/Fe]$=-0.44$). The Mg abundance is
determined from the \ion{Mg}{1} b lines, which are rather sensitive to the
adopted broadening parameter. However, the non-detection of other \ion{Mg}{1}
lines (e.g., 5528~{\AA}) results in an upper limit of [Mg/Fe]$\sim 0.0$,
indicating a deficiency of Mg in this star. The [Ca/Fe] ratio of this star is
also below the solar value. Moreover, the carbon abundance of this object is
very low ([C/Fe] $= -0.7$), as mentioned in \S~\ref{sec:ana_c}. The Na is also
significantly under-abundant ([Na/Fe] $= -1.0$). Further detailed abundance
study of this object is desirable, as a candidate VMP star revealing a peculiar
nucleosynthesis episode in the early chemical enrichment of the Galaxy.

Another object (SDSS J0254+3328: [Mg/Fe] $=-0.3$) also exhibits a relatively low
Ca abundance ([Ca/Fe] $=0.0$) for a star at this very low metallicity ([Fe/H] $
=-2.8$), and could be an $\alpha$-element-deficient star. The other two stars,
SDSS J1241$-$0837 and SDSS J1633+3907, have normal Ca abundances for halo stars,
and the deficiency of the $\alpha$-elements in general is unclear.

Excluding such outliers, no clear scatter of the [$\alpha$/Fe] ratios is
detected, within the measurement errors, in our sample. This indicates that
there is no large scatter in the abundance ratios of these elements even at very
low metallicity, as suggested by previous studies
\citep{francois04, arnone05, andrievsky10}. Higher quality spectra are
required to investigate the small size of the abundance scatter, if any, which
will provide useful constraints on the early chemical enrichment of the
Galaxy by supernovae and subsequent mixing in the interstellar medium.

\subsubsection{Abundance Trends}

The averages of [Mg/Fe] and [Ca/Fe] clearly exhibit over-abundances of
these elements in EMP stars, as have been found by numerous previous
studies \citep[e.g.,][]{ryan96, mcwilliam97, cayrel04}. There is no
clear increasing or decreasing trend of the [Mg/Fe] and [Ca/Fe]
abundance ratios in the sample taken from the SAGA database, as shown
by the dashed lines in Figure~\ref{fig:mgca}. This is also the case for
stars with [Fe/H]$ \leq -2.8$ in our sample; the average values of
[Mg/Fe] and [Ca/Fe] are +0.4 and +0.3, respectively, in agreement with
those reported by many previous studies \citep[e.g., ][]{lai08,
  andrievsky10}.

However, the abundance ratios at [Fe/H] $=-2.6$ ($\langle$[Mg/Fe]$\rangle=+0.08$
and $\langle$[Ca/Fe]$\rangle = -0.04$) are lower than these averages for the
stars with [Fe/H] $<-2.8$. Although the standard deviations are as large as
0.25~dex, the difference is statistically significant, given the sample size (11
objects) in this bin. Indeed, the null hypothesis that the [Mg/Fe] for stars in
the bin ($-2.7<$ [Fe/H] $\leq-2.5$) and for the lower metallicity stars ([Fe/H]
$\leq -2.7$) are drawn from the same parent population is rejected by the
Mann-Whitney rank-sum test at high significance ($p < 0.001$).

This metallicity bin includes a relatively large fraction of giants
(six giants among the 11 stars). However, there is no significant
difference in the average [Mg/Fe] between giants and turn-off stars in
our entire sample (the difference is less than 0.01~dex). Moreover,
the average [Mg/Fe] for the six giants in the metallicity range $-2.7
< $[Fe/H]$ \leq -2.5$ ($<$[Mg/Fe]$>$ = $+0.16$) is rather higher than
for the five turn-off stars in the bin ($<$[Mg/Fe]$>$ =
$-0.03$). Hence, the relatively large fraction of giants in this bin
is unlikely to be the reason for the low [Mg/Fe].

The [Mg/Ca] ratio is almost constant over the full metallicity range in
our sample, as found by previous studies. Hence, the
abundance trend of our sample suggests a decreasing trend of the
$\alpha$-elements at [Fe/H]$\sim -2.5$.

Recent abundance studies of halo stars with available full space motions suggest
different trends in the [$\alpha$/Fe] abundance ratios depending on kinematics
\citep{zhang09, ishigaki10, nissen10}. These studies mostly include less
metal-poor stars ([Fe/H]$\gtrsim -2.0$) than our sample. Further investigations
of the kinematics, as well as their chemical abundance ratios, for significantly more
metal-poor stars, such as those included in our present sample, is desired to
understand the early formation processes of the Galactic halo system.

\subsection{Sr and Ba}

\begin{figure}
\epsscale{.42}
\plotone{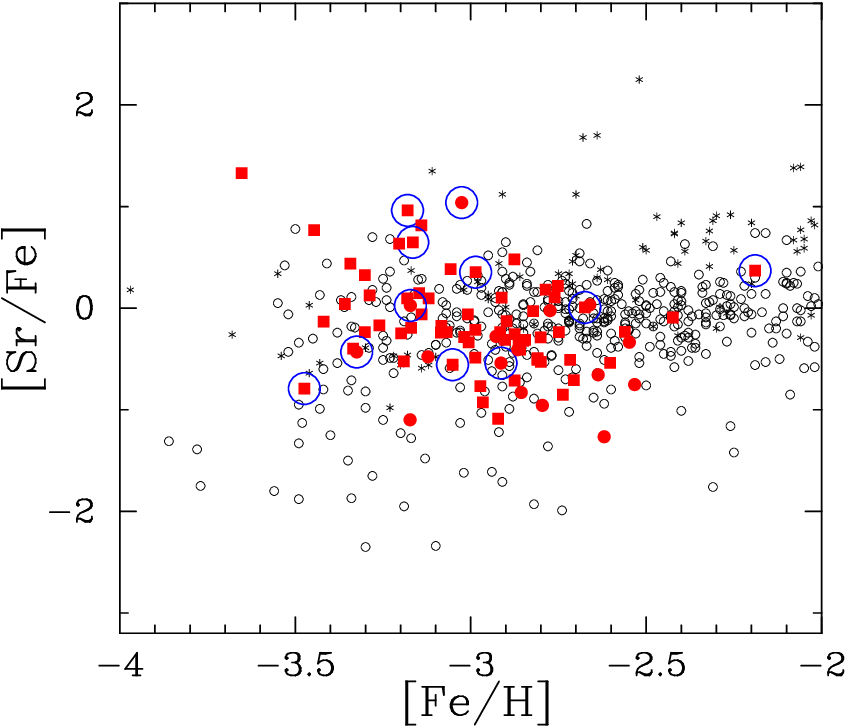} \\
\plotone{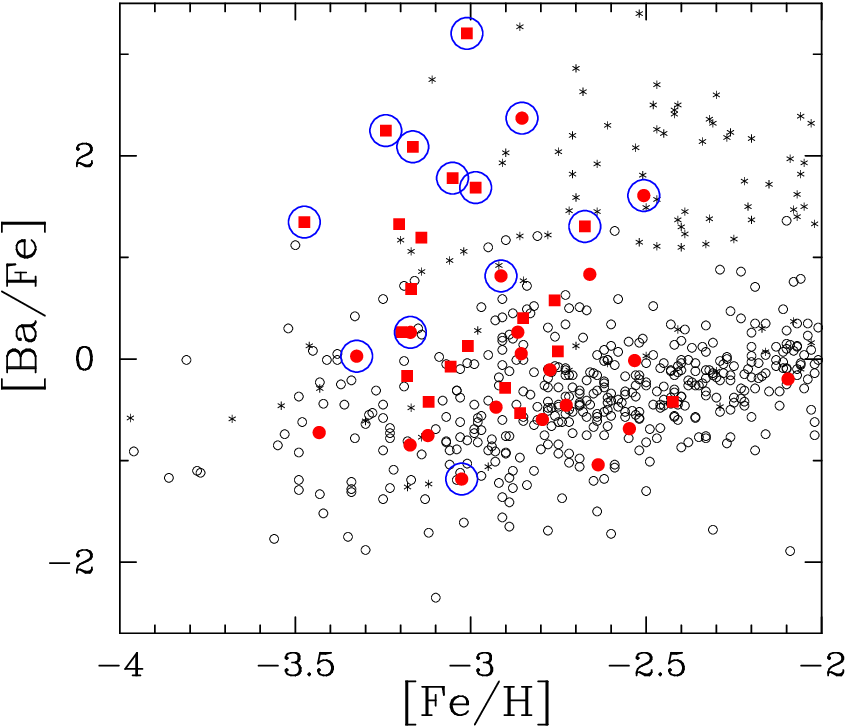} \\
\plotone{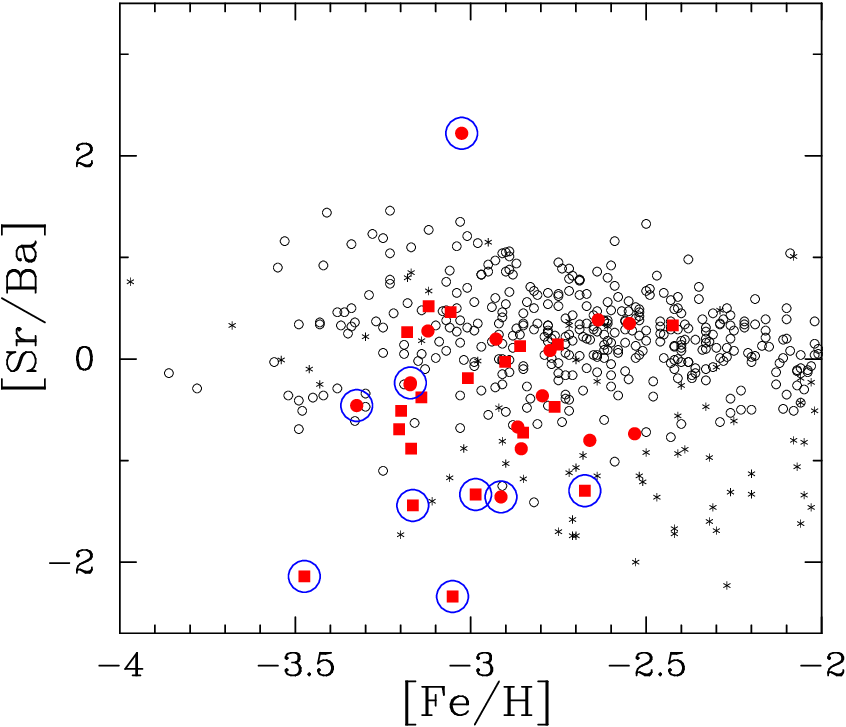}

\caption{Abundance ratios of neutron-capture elements ([Sr/Fe],
  [Ba/Fe], and [Sr/Ba]) as a function of [Fe/H].  The filled squares
  and circles (red) indicate main-sequence turn-off and giant stars,
  respectively. CEMP stars ([C/Fe] $> +0.7$) are over-plotted by large
open circles (blue). Small open circles and asterisks indicate the
abundance ratios of carbon-normal and carbon-rich stars, respectively,
measured by previous studies, which are taken from the SAGA
database.  \label{fig:srba}}

\end{figure}

Figure~\ref{fig:srba} shows the abundance ratios of [Sr/Fe], [Ba/Fe],
and [Sr/Ba] as a function of metallicity. The results are compared
with those obtained by previous studies, also taken from the SAGA
database \citep{suda08}, which are shown by open circles
(carbon-normal stars) and asterisks (carbon-enhanced stars).

Non-LTE effects on Sr and Ba abundance determinations were
  investigated by \citet{andrievsky11} and \citet{andrievsky09},
  respectively. The abundance corrections for Ba are positive, hence
  our LTE analyses could systematically underestimate the Ba
  abundances, though the correction would be at most 0.3~dex. The
  correction for Sr abundances determined from the \ion{Sr}{2} 4077
  and 4215~{\AA} lines could be positive and negative, depending on
  the stellar parameters. The corrections are, however, at most
  0.2~dex. Since these effects are much smaller than the scatter found
  in the Sr and Ba abundance ratios, our discussion here is not
  significantly affected by the non-LTE effects.

The [Sr/Fe] ratios (top panel) exhibit a scatter of about one order of
magnitude. Interestingly, this scatter is much {\it smaller} than that
found by previous studies at [Fe/H] $\sim -3$
\citep[e.g., ][]{mcwilliam95, ryan96, honda04, aoki05,
  francois07}. However, this could be the result of a bias in the
sample, since the \ion{Sr}{2} lines are in the blue range (where the
spectral data quality is not high), and are only detected in stars
having high Sr abundance. Indeed, the objects in our sample distribute
in the range of [Sr/Fe]$\gtrsim -1$, below which many stars are found
in the SAGA sample. In other words, there are likely to be many stars
having lower [Sr/Fe] ratios in our sample for which the Sr lines are
not detected.

Although the situation is similar for the [Ba/Fe] ratios (middle
panel), the scatter is much larger than for [Sr/Fe]. This is mostly
due to the large excesses of Ba in carbon-enhanced stars (the CEMP-s
stars), which are shown by over-plotting large open circles.  This is
clear from the comparison with previous studies: carbon-enhanced
objects in the SAGA sample are shown by asterisks in
Figure~\ref{fig:srba}.
The s-process at low metallicity is known to yield larger amounts of heavy
neutron-capture elements, such as Ba, compared to lighter elements
such as Sr \citep[e.g., ][]{busso99, bisterzo11}.  This is clearly seen
in the [Sr/Ba] ratios (bottom panel), where most of the CEMP stars
exhibit low [Sr/Ba] ratios. There is one exception, at [Fe/H]$ =
-3.0$, that has a very high [Sr/Ba] ratio ([Sr/Ba] $=+2.2$). This
star, SDSS J1422+0031, exhibits no excess of Ba ([Ba/Fe] $ = -1.0$),
and is classified as a CEMP-no star.
                                   
Excluding the CEMP-s stars, four other stars have [Ba/Fe]
$>+0.5$. Among them, SDSS J2357--0052 is a highly r-process-enhanced
(r-II) star, reported on in detail by \citet{aoki10}. This object is
the first example of a cool EMP main-sequence star with large excesses
of r-process elements. The metallicity is the lowest, and the excess
of Eu  is the highest ([Eu/Fe]=+1.9), among the r-II stars known to date. We
note that the Fe abundance of this object derived in the present work
([Fe/H]$ =-3.2$) is slightly higher than the result of \citet{aoki10},
because the {\teff} adopted here is slightly higher.

SDSS J0932+0241 is another EMP star exhibiting a large excess of Ba ([Ba/Fe]$
=+1.2$). Because of the limited quality of our spectrum and the star's high temperature
({\teff}$=6200$~K), the abundances of most other heavy elements are not
determined. We note that the [Sr/Ba] ratio of this star ([Sr/Ba] $ =-0.3$) is
significantly higher than the values found in CEMP-s stars, suggesting the
origin of these heavy elements are attributable to the r-process, rather than to the
s-process. If this is confirmed, this object is the first clear example of r-II
stars at the main-sequence turn-off \citep{sneden08}. Further detailed abundance
study is desirable for this object to firmly establish the origin of the excess
in Ba.

The other two stars, SDSS J0008--0529 and SDSS J2128--0756, exhibit
[Ba/Fe] ratios of +0.6 and +0.8, respectively. If the origin of the Ba
in these stars is the r-process, the [Eu/Fe] values are expected to be
higher than +1.  Measurements of the heavy elements in these objects,
based on higher quality spectra, are also desirable for further
studies of r-II stars.

Another interesting object is SDSS J0140+2344, which has a large
overabundance of Sr ([Sr/Fe]$>+1$) with no clear excess of Ba. Though
many metal-poor stars having high Sr/Ba ratios are known \citep[e.g.,
][]{honda04, francois07}, this object is unique because of its low
metallicity ([Fe/H]$=-3.7$). Further detailed abundance study is
desired to understand the implication of the Sr overabundance in this
object.

\section{Summary}

We have determined stellar parameters and chemical compositions, based on
high-resolution spectra obtained with the Subaru/HDS, for
137 very/extremely metal-poor stars selected from SDSS/SEGUE. Comparisons of
the Fe abundances derived by the present work with the estimates by
the recent pipeline analyses for the SDSS spectra (SSPP) exhibit no
significant offset, even in the lowest metallicity range ([Fe/H]$<-3$),
while scatter in the comparisons indicates that high-resolution
spectroscopy is required to determine accurate metallicity for
individual stars. The abundance ratios of carbon, the $\alpha$-elements,
and the neutron-capture elements derived from our high-resolution spectra
will provide useful calibrations for the estimates from SDSS spectra. 

The fraction of carbon-enhanced objects and the abundance ratios of
$\alpha$-elements and neutron-capture elements are discussed for the
overall sample. More detailed abundance patterns will be studied based
on higher-resolution, higher-S/N spectra for selected objects, in
particular those having the lowest metallicity ([Fe/H]$\lesssim -3.5$), .

Our sample includes three double-lined spectroscopic binaries
(including a triple system), for which chemical compositions of the
primary stars are estimated taking the veiling by the secondary into
consideration. Follow-up studies for these binaries will be useful for
understanding low-mass star formation at low metallicity in the
early era of the Galaxy.

\acknowledgments

Funding for the SDSS and SDSS-II has been provided by the Alfred
P. Sloan Foundation, the Participating Institutions, the National
Science Foundation, the U.S. Department of Energy, the National
Aeronautics and Space Administration, the Japanese Monbukagakusho, the
Max Planck Society, and the Higher Education Funding Council for
England. The SDSS Web site is http://www.sdss.org/.

The SDSS is managed by the Astrophysical Research Consortium for the
Participating Institutions. The Participating Institutions are the
American Museum of Natural History, Astrophysical Institute Potsdam,
University of Basel, Cambridge University, Case Western Reserve
University, University of Chicago, Drexel University, Fermilab, the
Institute for Advanced Study, the Japan Participation Group, Johns
Hopkins University, the Joint Institute for Nuclear Astrophysics, the
Kavli Institute for Particle Astrophysics and Cosmology, the Korean
Scientist Group, the Chinese Academy of Sciences (LAMOST), Los Alamos
National Laboratory, the Max-Planck-Institute for Astronomy (MPIA),
the Max-Planck-Institute for Astrophysics (MPA), New Mexico State
University, Ohio State University, University of Pittsburgh,
University of Portsmouth, Princeton University, the United States
Naval Observatory, and the University of Washington.

W.A. was supported by the JSPS Grants-in-Aid for Scientific Research
(23224004).  T.C.B. and Y.S.L. acknowledge partial funding of this
work from grants PHY 02-16783 and PHY 08-22648: Physics Frontier
Center/Joint Institute for Nuclear Astrophysics (JINA), awarded by the
U.S. National Science Foundation. M.T.-H. is grateful for a support by
the JSPS Grants-in-Aid for Scientific Research (22540255).



{\it Facilities:} \facility{SDSS}, \facility{Subaru (HDS)}



\appendix
\section{Objects Not Analyzed}

\begin{figure}
\epsscale{.50}
\plotone{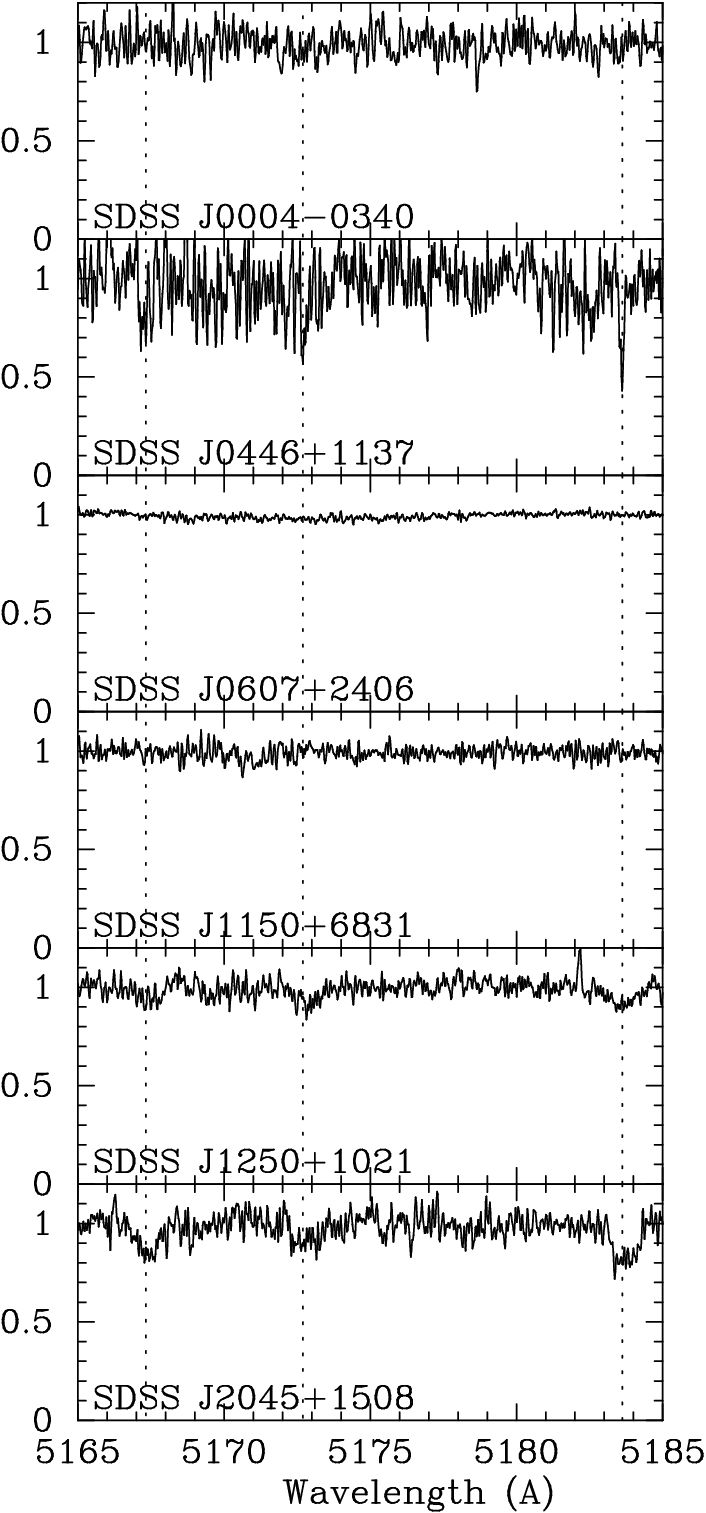}
\caption{Spectra of objects not analyzed in the present work, in the
  region of the \ion{Mg}{1} b lines. The line positions of the triplet
  are shown by vertical dotted lines. \label{fig:spnon1}}
\end{figure}

\begin{figure}
\epsscale{.50}
\plotone{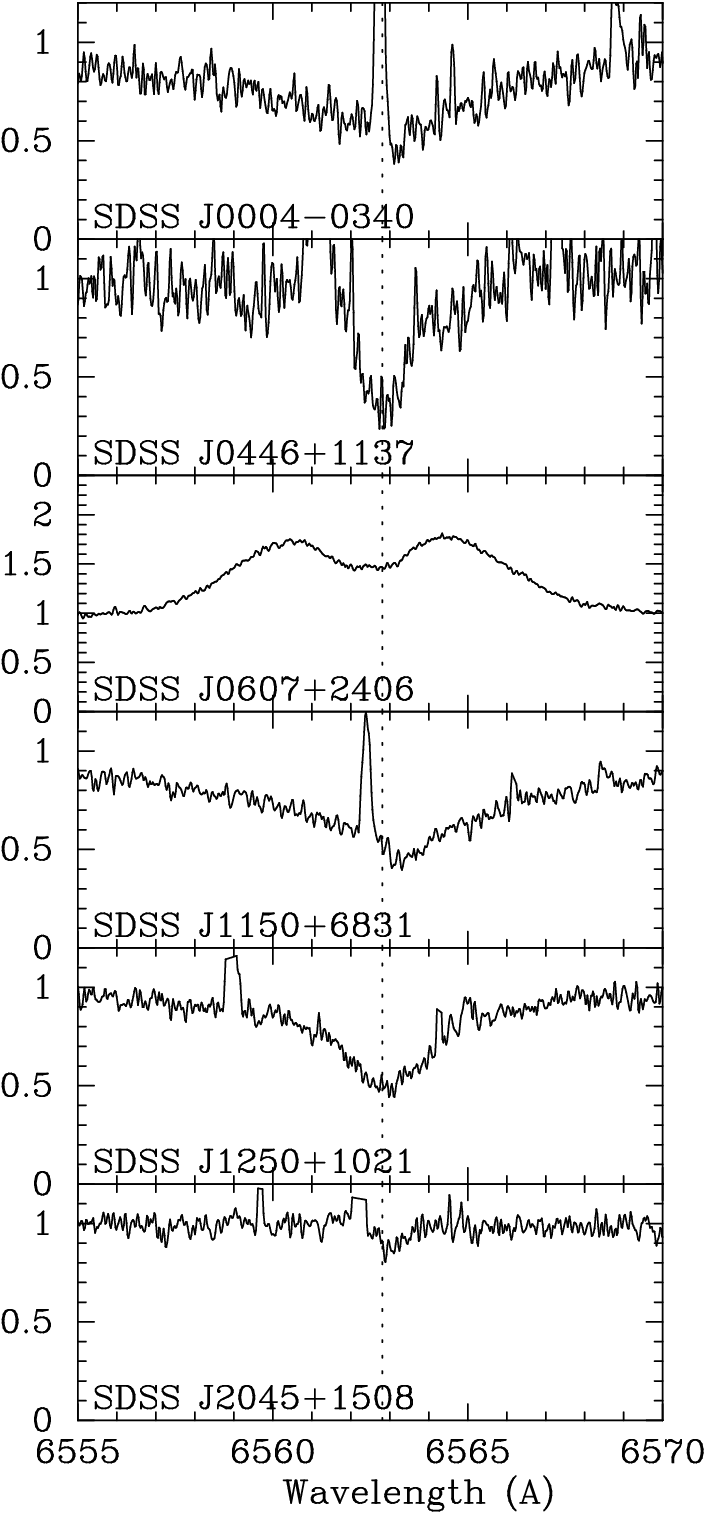}
\caption{The same as Fig.~\ref{fig:spnon1}, but for the H$\alpha$
  region. The line position is shown by a vertical dotted line. Sky
  emission features are not fully removed in this
  region. \label{fig:spnon2}}
\end{figure}

The six objects observed, but not analyzed, in the present work are listed
in Table~\ref{tab:objnon}. The spectra around the wavelengths of
\ion{Mg}{1} b lines and H$\alpha$ are shown in Figures~\ref{fig:spnon1}
and \ref{fig:spnon2}. 

Two objects (SDSS J0004$-$0340 and SDSS J1150$+$6831) are included in
the list of white dwarfs reported by \citet{debes11}. They exhibit broad
and shallow H$\alpha$ absorption lines and no clear \ion{Mg}{1} b lines
(Fig.~\ref{fig:spnon1} and \ref{fig:spnon2}). SDSS J1250$+$1021 and
SDSS J2045$+$1508 also show broad or shallow H$\alpha$ absorption
features with weak \ion{Mg}{1} b lines. They could be relatively cool
white dwarfs, though further confirmation is required.

SDSS J0607$+$2406 is a bright object, but exhibits an emission feature of
H$\alpha$, and is clearly not a normal metal-poor star. This object is
close to the cluster NGC~2168, and identification with objects
reported by previous studies is not straightforward.

SDSS J0446+1137 is likely a metal-poor star, though the S/N ratio of
the current spectrum is not sufficient for abundance analyses.

\clearpage


\begin{deluxetable}{llllccccc}
\tabletypesize{\scriptsize}
\tablewidth{0pt}
\tablecaption{OBJECT LIST \label{tab:objlist}}
\tablehead{
\colhead{ID} & \colhead{Object} & \colhead{Object name} & \colhead{Object ID} & \colhead{$g_{0}$} & \colhead{$(g-r)_{0}$} & \colhead{$V_{0}$}\tablenotemark{a} & \colhead{$E(B-V)$}\tablenotemark{b} & \colhead{$K$} \tablenotemark{c}
}
\startdata
001 & SDSS J0002$+$2928 & SDSS J000219.87$+$292851.8 & 2803-54368-459 & 15.133 & 0.304 & 14.958 & 0.051 & 13.683 \\
002 & SDSS J0008$-$0529 & SDSS J000812.54$-$052926.5 & 2624-54380-061 & 16.564 & 0.737 & 16.147 & 0.033 & 13.620 \\
003 & SDSS J0018$-$0939 & SDSS J001820.51$-$093939.2 & 1912-53293-352 & 16.140 & 0.789 & 15.693 & 0.045 & 13.278 \\
004 & SDSS J0020$-$0040 & SDSS J002015.45$-$004058.1 & 1121-52873-136 & 16.619 & 0.221 & 16.491 & 0.028 & 15.069 \\
005 & SDSS J0021$-$0050 & SDSS J002113.78$-$005005.2 & 0390-51816-187 & 16.571 & 0.262 & 16.420 & 0.028 & 15.403 \\
006 & SDSS J0023$-$0003 & SDSS J002356.26$-$000311.9 & 0688-52203-152 & 16.454 & 0.405 & 16.223 & 0.028 & 14.505 \\
007 & SDSS J0027$+$1404 & SDSS J002749.46$+$140418.1 & 0753-52233-013 & 16.533 & 0.278 & 16.373 & 0.102 & 15.013 \\
008 & SDSS J0027$-$1909 & SDSS J002756.76$-$190929.8 & 2848-54453-300 & 15.574 & 0.204 & 15.456 & 0.024 & 14.390 \\
\enddata
\tablenotetext{a}{$V_{0}$ is derived by the transform of \citet{zhao06}.}
\tablenotetext{b}{$E(B-V)$ is estimated from \citet{schlegel98}.}
\tablenotetext{c}{$K$ is taken from 2MASS \citep{skrutskie06}.}
\end{deluxetable}

\begin{deluxetable}{llllcccrrr}
\tabletypesize{\scriptsize}
\tablewidth{0pt}
\tablecaption{OBSERVING LOG \label{tab:obslog}}
\tablehead{
\colhead{ID}  & \colhead{Object}          & \colhead{Obs.date(1)} & \colhead{Obs.date(2)} & \colhead{exp.time} & \colhead{$S/N$} & \colhead{$S/N$} & \colhead{$V_{\rm H}$} & \colhead{$\sigma(V_{\rm H})$} & \colhead{$V_{\rm H}^{\rm SDSS}$} \\
\colhead{  }  & \colhead{     }          & \colhead{            } & \colhead{           } & \colhead{(minute)} & \colhead{4300{\AA}} & \colhead{5180{\AA}} & \colhead{({\kms})        } & \colhead{({\kms})               } & \colhead{({\kms})}
}
\startdata
001 & SDSS J0002$+$2928 & July 6, 2008    &  & 20 & 40 & 65            &$  -308.22$&$  0.18$&$    -297.2$\\
002 & SDSS J0008$-$0529 & October 4, 2008 &  & 25 & 24 & 46            &$   117.18$&$  0.07$&$     117.5$\\
003 & SDSS J0018$-$0939 & July 5, 2008    &  & 20 & 26 & 51            &$  -122.89$&$  0.07$&$    -119.9$\\
004 & SDSS J0020$-$0040 & October 5, 2008 &  & 40 & 35 & 54            &$     3.29$&$  0.13$&$      11.2$\\
005 & SDSS J0021$-$0050 & July 5, 2008    &  & 30 & 32 & 49            &$   -94.57$&$  0.19$&$     -90.2$\\
006 & SDSS J0023$-$0003 & October 4, 2008 &  & 25 & 28 & 46            &$    89.84$&$  0.10$&$      88.0$\\
007 & SDSS J0027$+$1404 & October 5, 2008 &  & 40 & 31 & 51            &$    27.74$&$  0.24$&$      35.9$\\
008 & SDSS J0027$-$1909 & July 5, 2008 & July 6, 2008 & 40 & 54 & 83   &$    14.97$&$  0.10$&$      15.0$\\
\enddata
\tablenotetext{a}{Objects which were later observed at higher resolution and higher S/N.}
\end{deluxetable}

\begin{deluxetable}{rcrrl}
\tabletypesize{\scriptsize}
\tablewidth{0pt}
\tablecaption{SPECTRAL LINE DATA \label{tab:lines}}
\tablehead{
\colhead{Species} & \colhead{Wavelength} & \colhead{$ \log gf$} & \colhead{L.E.P.} & \colhead{References\tablenotemark{a}} \\
 & \colhead{(\AA)} & & \colhead{(eV)} &  
}
\startdata
   O I  &   6300.304 &     $-$9.819 &    0.000 & 1        \\
  Na I  &   5889.951 &      0.117 &    0.000 & 1        \\
  Na I  &   5895.924 &     $-$0.184 &    0.000 & 1         \\
  Mg I  &   4167.271 &     $-$0.710 &    4.346 & 2  \\
  Mg I  &   4571.096 &     $-$5.393 &    0.000 & 3          \\
  Mg I  &   4702.991 &     $-$0.380 &    4.330 & 2           \\
  Mg I  &   5172.684 &     $-$0.450 &    2.712 & 4  \\
  Mg I  &   5183.604 &     $-$0.239 &    2.717 & 4  \\
\enddata
\tablenotetext{a}{References -- (1)\citet{kupka99}; (2) \citet{fischer75}; (3)\citet{wiese69}; (4)\citet{aldenius07}; (5)\citet{wiese80}; (6)\citet{ivans06}; (7)\citet{lawler89}; (8)\citet{blackwell82a}; (9)\citet{blackwell82b}; (10)\citet{pickering01}; (11)\citet{ryabchikova94}; (12)\citet{martin88}; (13)\citet{booth84}; (14)\citet{obrian91}; (15) \citet{schnabel04}; (16)\citet{moity83}; (17); \citet{biemont91}; (18) \citet{nitz99}; (19) \citet{biemont80}; (20) \citet{pinnington95}; (21) \citet{hannaford82}; (22) \citet{gallagher67}; (23)\citet{lawler01la}; (24)\citet{lawler01eu}}
\end{deluxetable}



\begin{deluxetable}{llccccccccc}
\tabletypesize{\scriptsize}
\tablewidth{0pt}
\tablecaption{STELLER PARAMETERS \label{tab:param}}
\tablehead{
 \colhead{ID} & \colhead{Object} & \multicolumn{3}{c}{SSPP} & & \colhead{{\teff}(K)} & \colhead{{\teff}(K)} & \multicolumn{3}{c}{Adopted parameters} \\
\cline{3-5} \cline{9-11} 
     &                 & \colhead{{\teff}(K)} & \colhead{$\log g$} & \colhead{[Fe/H]} && \colhead{$(V-K)_{0}$} & \colhead{$(g-r)_{0}$} & \colhead{{\teff}(K)} & \colhead{$\log g$} & \colhead{[Fe/H]} 
} 
\startdata
001 & SDSS J0002$+$2928 & 6169 & 3.10 & $-$2.81 &  & 6328 & 6043 & 6150 & 4.0 & $-$3.26 \\
002 & SDSS J0008$-$0529 & 4938 & 2.51 & $-$2.78 &  & 4549 & 4673 & 4950 & 2.0 & $-$2.66 \\
003 & SDSS J0018$-$0939 & 4612 & 3.29 & $-$2.98 &  & 4648 & 4628 & 4600 & 5.0 & $-$2.65 \\
004 & SDSS J0020$-$0040 & 6453 & 2.69 & $-$2.61 &  & 6055 & 6444 & 6450 & 4.0 & $-$2.81 \\
005 & SDSS J0021$-$0050 & 6654 & 3.31 & $-$3.09 &  & 6944 & 6242 & 6650 & 4.0 & $-$3.16 \\
006 & SDSS J0023$-$0003 & 5499 & 2.18 & $-$3.04 &  & 5555 & 5618 & 5500 & 4.0 & $-$3.12 \\
007 & SDSS J0027$+$1404 & 6270 & 3.68 & $-$2.88 &  & 6121 & 6166 & 6250 & 4.0 & $-$3.03 \\
008 & SDSS J0027$-$1909 & 6555 & 3.80 & $-$3.04 &  & 6824 & 6529 & 6550 & 4.0 & $-$2.71 \\
\enddata
\end{deluxetable}

\begin{deluxetable}{cccccccccc}
\tabletypesize{\scriptsize}
\tablewidth{0pt}
\tablecaption{CHEMICAL ABUNDANCES RESULTS \label{tab:abundance}}
\tablehead{
 & \colhead{[Fe/H](\ion{Fe}{1})} & \colhead{[Fe/H](\ion{Fe}{2})} & \colhead{[Na/Fe]} & \colhead{[Mg/Fe]} & \colhead{[Ca/Fe]} & \colhead{[Ti/Fe]} & \colhead{[Sr/Fe]} & \colhead{[Ba/Fe]} & \colhead{[C/Fe]}
}
\startdata
001 & \multicolumn{9}{c}{SDSS J0002+2928} \\ 
 \tableline 
 $\log \epsilon$   &  4.24 &  4.59 &  3.97 &  4.70 &  2.87 &  2.49 & $-$0.12 &  0.75 &  7.80 \\           
 $[$X/Fe$]$  & $-$3.26 &  0.35 &  1.00 &  0.36 & $-$0.20 &  0.80 &  0.27 &  1.84 &  2.63 \\       
 $\sigma_{\rm tot}$ &  0.21 &  0.30 &  0.24 &  0.16 &  0.21 &  0.27 &  0.29 &  0.27 &  0.22 \\
 $N$       &    11 &     2 &     1 &     2 &     1 &     4 &     1 &     2 & \nodata \\           
 \tableline 
002 & \multicolumn{9}{c}{SDSS J0008$-$0529} \\ 
 \tableline 
 $\log \epsilon$   &  4.84 &  4.86 &  3.79 &  5.18 &  3.77 &  2.61 &  0.24 &  0.35 &\nodata\\           
 $[$X/Fe$]$  & $-$2.66 &  0.02 &  0.21 &  0.24 &  0.09 &  0.32 &  0.03 &  0.83 &\nodata\\
\enddata
\end{deluxetable}

\begin{deluxetable}{lrrrrrcrrrrr}
\tabletypesize{\scriptsize}
\tablewidth{0pt}
\tablecaption{ABUNDANCE CHANGES BY CHANGING ATMOSPHERIC PARAMETERS \label{tab:error}}
\tablehead{
 & \multicolumn{5}{c}{giant} & & \multicolumn{5}{c}{turn-off} \\
\cline{2-6}  \cline{8-12}  
\colhead{SPECIES} & \colhead{$\delta${\teff}} & \colhead{$\delta \log g$} & \colhead{$\delta$[Fe/H]} & \colhead{$\delta${\vt}} & \colhead{r.s.s.} &  & \colhead{$\delta${\teff}} & \colhead{$\delta \log g$} & \colhead{$\delta$[Fe/H]} & \colhead{$\delta${\vt}} & \colhead{r.s.s} 
}
\startdata
 Fe (Fe I) & 0.19 & $-0.10$ & 0.00 & -0.26 & 0.33 & & 0.13 & -0.04 & 0.01 & 0.14 & 0.20 \\
 Na & $-0.03$  &  0.03 &  0.00 & $-0.12$ & 0.13  &  & $-$0.03 & 0.03 & 0.00 & $-$0.12 & 0.13 \\
 Mg & $-$0.01 & $-$0.10 & $-$0.01 & 0.03 & 0.10  & & $-$0.02 & $-$0.05 & 0.00 & $-$0.02 & 0.06 \\
 Ca & 0.05   & $-$0.14 & 0.00 & 0.03 & 0.15  &  & $-$0.02 & $-$0.01 & 0.00 & $-$0.04 & 0.04 \\
 Ti (Ti II) & $-$0.11 & 0.24 & 0.00 & 0.13 & 0.30 &  & $-$0.07 & 0.21 & $-$0.01 & $-$0.12 & 0.25 \\
 Fe (Fe II) & $-$0.16 & 0.25 & 0.00 & 0.13 & 0.32 &  & $-$0.11 & 0.21 & $-$0.01 & $-$0.12 & 0.26 \\
 Sr & $-0.05$ & 0.19 & 0.00 & $-0.04$ & 0.20   & & $-$0.05 & 0.19 & 0.00 & $-$0.04 & 0.20 \\
 Ba & $-$0.07 & 0.25 & 0.00 & 0.22 & 0.34  & & $-$0.04 & 0.20 & 0.00 & $-$0.11 & 0.23 \\
\enddata
\end{deluxetable}

\begin{deluxetable}{llrrcrr}
\tablecaption{CHEMICAL ABUNDANCES OF G~64--12 \label{tab:comparisonstar}}
\tablewidth{0pt}
\tablehead{
\colhead{Element} & \colhead{Species} &  \multicolumn{2}{c}{This work} & & \multicolumn{2}{c}{Aoki et al. (2006)} \\
\cline{3-4} \cline{6-7} 
       &  &  \colhead{$\log\epsilon$} & \colhead{[X/Fe]} && \colhead{$\log\epsilon$} & \colhead{[X/Fe]}}
\startdata
Fe & \ion{Fe}{1} & 4.36 & $-3.14$ && 4.25 & $-3.25$ \\
Fe & \ion{Fe}{2} & 4.24 & $-3.26$ && 4.38 & $-3.12$ \\
Na & \ion{Na}{1} & 2.82 & $-0.28$ && 2.74\tablenotemark{a} & \nodata \\
Mg & \ion{Mg}{1} & 4.79 & $0.33$  && 4.80 & $0.45$ \\
Ca & \ion{Ca}{1} & 3.50 & $0.30$  && 3.62 & $0.53$ \\
Ti & \ion{Ti}{2} & 2.27 & $0.46$  && 2.20 & $0.50$ \\
Sr & \ion{Sr}{2} & $-0.35$ & $-0.08$ && $-0.10$ & $0.28$ \\
\enddata
\tablenotetext{a}{Value adopted from Aoki et al. (2009)}
\end{deluxetable}

\begin{deluxetable}{llllcccl}
\tabletypesize{\scriptsize}
\tablewidth{0pt}
\tablecaption{LIST OF OBJECTS NOT ANALYZED \label{tab:objnon}}
\tablehead{
\colhead{ID} & \colhead{Object} & \colhead{Object name} & \colhead{Object ID} & 
\colhead{$g_{0}$} & \colhead{$(g-r)_{0}$} & \colhead{$V_{0}$} &
 \colhead{Remarks} 
}
\startdata
U1 & SDSS J0004$-$0340 & SDSS J000410.42$-$034008.6 & 2624-54380-458 & 16.762& 0.131& 16.685 & LP~644--30, WD\tablenotemark{a} \\
U2 & SDSS J0446$+$1137 & SDSS J044655.70$+$113741.3 & 2669-54086-593 & 16.376 & 0.248 & 16.232 & \\
U3 & SDSS J0607$+$2406 & SDSS J060740.48$+$240651.3 & 2887-54521-537 & 13.690 &  0.470 & 13.422 & close to NGC~2168\\
U4 & SDSS J1150$+$6831 & SDSS J115052.32$+$683116.1 & 0492-51955-523 & 15.337& 0.230& 15.204 & WD\tablenotemark{a} \\
U5 & SDSS J1250$+$1021 & SDSS J125005.10$+$102156.4 & 2963-54589-474 & 16.276& 0.257& 16.128 & \\
U6 & SDSS J2045$+$1508 & SDSS J204524.04$+$150825.5 & 2250-53566-249 & 16.358& 0.441& 16.107 & \\
\enddata
\tablenotetext{a}{White dwarfs listed by \citet{debes11}.}
\end{deluxetable}


\begin{thebibliography}{}

\bibitem[Abazajian et al.(2009)]{abazajian09} Abazajian, K.~N., Adelman-McCarthy, J.~K., Ag{\"u}eros, M.~A., et al.\ 2009, \apjs, 182, 543 

\bibitem[Aldenius et 
al.(2007)]{aldenius07} Aldenius, M., Tanner, J.~D., Johansson, S., Lundberg, H., \& Ryan, S.~G.\ 2007, \aap, 461, 767 


\bibitem[Allende Prieto et al.(2008)]{allendeprieto08} Allende Prieto, C., Sivarani, T., Beers, T.~C., et al.\ 2008, \aj, 136, 2070 

\bibitem[Alonso et al.(1999)]{alonso99} Alonso, A., Arribas, S., \& Mart{\'{\i}}nez-Roger, C.\ 1999, \aaps, 140, 261 


\bibitem[Andrievsky et al.(2009)]{andrievsky09} Andrievsky, S.~M., Spite, M., Korotin, S.~A., et al.\ 2009, \aap, 494, 1083 

\bibitem[Andrievsky et al.(2010)]{andrievsky10} Andrievsky, S.~M., Spite, M., Korotin, S.~A., et al.\ 2010, \aap, 509, A88 

\bibitem[Andrievsky et al.(2011)]{andrievsky11} Andrievsky, S.~M., Spite, F., Korotin, S.~A., et al.\ 2011, \aap, 530, A105 

\bibitem[Aoki et al.(2009)]{aoki09} Aoki, W., Barklem, P.~S., Beers, T.~C., et al.\ 2009, \apj, 698, 1803 

\bibitem[Aoki et al.(2007)]{aoki07} Aoki, W., Beers, T.~C., Christlieb, N., et al.\ 2007, \apj, 655, 492 

\bibitem[Aoki et al.(2010)]{aoki10} Aoki, W., Beers, T.~C., Honda, S., \& Carollo, D.\ 2010, \apjl, 723, L201 

\bibitem[Aoki et al.(2008)]{aoki08} Aoki, W., Beers, T.~C., Sivarani, T., et al.\ 2008, \apj, 678, 1351 

\bibitem[Aoki et al.(2006)]{aoki06} Aoki, W., Frebel, A., Christlieb, N., et al.\ 2006, \apj, 639, 897 

\bibitem[Aoki et al.(2005)]{aoki05} Aoki, W., Honda, S., Beers, T.~C. et al.\ 2005, \apj, 632, 611 

\bibitem[Aoki et al.(2002a)]{aoki02a} Aoki, W., Norris, J.~E., Ryan, S.~G., Beers, T.~C., \& Ando, H.\ 2002a, \pasj, 54, 427 

\bibitem[Aoki et al.(2002b)]{aoki02b} Aoki, W., Norris, J.~E., Ryan, S.~G., Beers, T.~C., \& Ando, H.\ 2002b, \apjl, 576, L141 

\bibitem[Arnone et al.(2005)]{arnone05} Arnone, E., Ryan, S.~G., Argast, D., Norris, J.~E., \& Beers, T.~C.\ 2005, \aap, 430, 507 


\bibitem[Asplund et al.(2009)]{asplund09} Asplund, M., Grevesse, N., Sauval, A.~J., \& Scott, P.\ 2009, \araa, 47, 481 

\bibitem[Barklem et al.(2005)]{barklem05} Barklem, P.~S., Christlieb, N., Beers, T.~C., et al.\ 2005, \aap, 439, 129 

\bibitem[Beers \& Christlieb (2005)]{beers05} Beers, T. C., \& Christlieb, N. 2005, ARAA, 43, 531 

\bibitem[Beers et al.(1985)]{beers85} Beers, T.~C., Preston, G.~W., \& Shectman, S.~A.\ 1985, \aj, 90, 2089 

\bibitem[Beers et al.(1992)]{beers92} Beers, T.~C., Preston, G.~W., \& Shectman, S.~A.\ 1992, \aj, 103, 1987 

\bibitem[Bessell \& Norris(1984)]{bessell84} Bessell, M.~S., \& Norris, J.\ 1984, \apj, 285, 622 

\bibitem[Biemont et 
al.(1991)]{biemont91} Biemont, E., Baudoux, M., Kurucz, R.~L., Ansbacher, W., \& Pinnington, E.~H.\ 1991, \aap, 249, 539 

\bibitem[Biemont 
\& Godefroid(1980)]{biemont80} Biemont, E., \& Godefroid, M.\ 1980, \aap, 84, 361 

\bibitem[Bisterzo et al.(2011)]{bisterzo11} Bisterzo, S., Gallino, R., Straniero, O., Cristallo, S., \& K\"{a}ppeler, F.\ 2011, \mnras, 418, 284 

\bibitem[Blackwell et al.(1982a)]{blackwell82a} Blackwell, D.~E., 
Menon, S.~L.~R., Petford, A.~D., \& Shallis, M.~J.\ 1982a, \mnras, 201, 611 

\bibitem[Blackwell et al.(1982b)]{blackwell82b} Blackwell, D.~E., 
Petford, A.~D., Shallis, M.~J., \& Leggett, S.\ 1982b, \mnras, 199, 21 

\bibitem[Bonifacio et al.(2012)]{bonifacio12} Bonifacio, P., Sbordone, L., Caffau, E., et al.\ 2012, \aap, 542, A87 

\bibitem[Bonifacio et 
al.(2009)]{bonifacio09} Bonifacio, P., Spite, M., Cayrel, R., et al.\ 2009, \aap, 501, 519 

\bibitem[Booth et al.(1984)]{booth84} Booth, A.~J., Blackwell, 
D.~E., Petford, A.~D., \& Shallis, M.~J.\ 1984, The Observatory, 104, 265 


\bibitem[Bromm 
\& Larson(2004)]{bromm04} Bromm, V., \& Larson, R.~B.\ 2004, \araa, 42, 79 

\bibitem[Busso et al.(1999)]{busso99} Busso, M., Gallino, R., \& Wasserburg, G.~J.\ 1999, \araa, 37, 239 

\bibitem[Caffau et al.(2011)]{caffau11} Caffau, E., Bonifacio, P., Fran{\c c}ois, P., et al.\ 2011, \nat, 477, 67 


\bibitem[Carollo et al.(2012)]{carollo12} Carollo, D., Beers, T.~C., Bovy, J., et al.\ 2012, \apj, 744, 195 

\bibitem[Casagrande et al.(2010)]{casagrande10} Casagrande, L., Ram{\'{\i}}rez, I., Mel{\'e}ndez, J., Bessell, M., \& Asplund, M.\ 2010, \aap, 512, A54 

\bibitem[Castelli \& Kurucz(2003)]{castelli03} Castelli, F., \& Kurucz, R.~L.\ 2003, Modelling of Stellar Atmospheres, 210, 20P

\bibitem[Cayrel et al.(2004)]{cayrel04} Cayrel, R., Depagne, E., Spite, M., et al.\ 2004, \aap, 416, 1117 

\bibitem[Christlieb(2003)]{christlieb03} Christlieb, N.\ 2003, Reviews in Modern Astronomy, 16, 191 

\bibitem[Christlieb et al.(2002)]{christlieb02} Christlieb, N., Bessell, M.~S., Beers, T.~C., et al.\ 2002, \nat, 419, 904 

\bibitem[Christlieb et al.(2008)]{christlieb08} Christlieb, N., Sch{\"o}rck, T., Frebel, A., et al.\ 2008, \aap, 484, 721 

\bibitem[Ciardi 
\& Ferrara(2005)]{ciardi05} Ciardi, B., \& Ferrara, A.\ 2005, \ssr, 116, 625 

\bibitem[Cohen et al.(2004)]{cohen04} Cohen, J.~G., Christlieb, N., McWilliam, A., et al.\ 2004, \apj, 612, 1107 


\bibitem[Cooke et al.(2011)]{cooke11} Cooke, R., Pettini, M., 
Steidel, C.~C., Rudie, G.~C., \& Jorgenson, R.~A.\ 2011, \mnras, 412, 1047 

\bibitem[Debes et al.(2011)]{debes11} Debes, J.~H., Hoard, 
D.~W., Wachter, S., Leisawitz, D.~T., \& Cohen, M.\ 2011, \apjs, 197, 38 

\bibitem[Demarque et al.(2004)]{demarque04} Demarque, P., Woo, J.-H., Kim, Y.-C., \& Yi, S.~K.\ 2004, \apjs, 155, 667


\bibitem[Fischer (1975)]{fischer75} Fischer, F. C. 1975, Canad. J. Phys. 53, 189


\bibitem[Fran{\c c}ois et al.(2007)]{francois07} Fran{\c c}ois, P., Depagne, E., Hill, V.,  et al.\ 2007, \aap, 476, 935 

\bibitem[Fran{\c c}ois et al.(2004)]{francois04} Fran{\c c}ois, P., Matteucci, F., Cayrel, R., et al.\ 2004, \aap, 421, 613 

\bibitem[Frebel et al.(2005)]{frebel05} Frebel, A., Aoki, W., Christlieb, N., et al.\ 2005, \nat, 434, 871 

\bibitem[Frebel \& Norris(2011)]{frebel11} Frebel, A., \& Norris, J.~E.\ 2011, arXiv:1102.1748



\bibitem[Gallagher(1967)]{gallagher67} Gallagher, A.\ 1967, 
Physical Review, 157, 24 


\bibitem[Goldberg et al.(2002)]{goldberg02} Goldberg, D., Mazeh, T., Latham, D.~W., et al.\ 2002, \aj, 124, 1132 


\bibitem[Gunn et al.(2006)]{gunn06} Gunn, J.~E., Siegmund, W.~A., Mannery, E.~J., et al.\ 2006, \aj, 131, 2332 

\bibitem[Hannaford et al.(1982)]{hannaford82} Hannaford, P., Lowe, 
R.~M., Grevesse, N., Biemont, E., \& Whaling, W.\ 1982, \apj, 261, 736 

\bibitem[Hirschi et al.(2006)]{hirschi06} Hirschi, R., Fr\"olich, C., Liebendorfer, M., \& Thilemann, F.-K.\ 2006, Reviews in Modern Astronomy, 19, 101

\bibitem[Hollek et al.(2011)]{hollek11} Hollek, J.~K., Frebel, A., Roederer, I.~U., et al.\ 2011, \apj, 742, 54 

\bibitem[Honda et al.(2004)]{honda04} Honda, S., Aoki, W., Kajino, T., et al.\ 2004, \apj, 607, 474
 
\bibitem[Ishigaki et al.(2010)]{ishigaki10} Ishigaki, M., Chiba, M., \& Aoki, W.\ 2010, \pasj, 62, 1369 

\bibitem[Ito et al.(2009)]{ito09} Ito, H., Aoki, W., Honda, S., \& Beers, T.~C.\ 2009, \apj, 698, L37

\bibitem[Ivans et al.(2006)]{ivans06} Ivans, I.~I., Simmerer, 
J., Sneden, C., et al.\ 2006, \apj, 645, 613 

\bibitem[Ivans et al.(2005)]{ivans05} Ivans, I.~I., Sneden, C., Gallino, R., Cowan, J.~J., \& Preston, G.~W.\ 2005, \apjl, 627, L145 

\bibitem[Kim et al.(2002)]{kim02} Kim, Y.-C., Demarque, P., Yi, S.~K., \& Alexander, D.~R.\ 2002, \apjs, 143, 499 

\bibitem[Kobayashi et al.(2011)]{kobayashi11} Kobayashi, C., Tominaga, N., \& Nomoto, K.\ 2011, \apjl, 730, L14 


\bibitem[Kupka et 
al.(1999)]{kupka99} Kupka, F., Piskunov, N., Ryabchikova, T.~A., Stempels, H.~C., \& Weiss, W.~W.\ 1999, \aaps, 138, 119 

\bibitem[Kurucz(1993)]{kurucz93} Kurucz, R.\ 1993, ATLAS9 Stellar Atmosphere Programs and 2 km/s grid.~Kurucz CD-ROM No.~13.~Cambridge, Mass.: Smithsonian Astrophysical Observatory, 1993, 13

\bibitem[Lai et al.(2008)]{lai08} Lai, D.~K., Bolte, M., Johnson, J.~A., et al.\ 2008, \apj, 681, 1524 

\bibitem[Lawler et al.(2001a)]{lawler01la} Lawler, J.~E., 
Bonvallet, G., \& Sneden, C.\ 2001a, \apj, 556, 452 

\bibitem[Lawler \& Dakin (1989)]{lawler89}
Lawler, J. E., \& Dakin, J. T. 1989, J. Opt. Soc. Amer., B6, 1457

\bibitem[Lawler et al.(2001b)]{lawler01eu} Lawler, J.~E., 
Wickliffe, M.~E., den Hartog, E.~A., \& Sneden, C.\ 2001b, \apj, 563, 1075 

\bibitem[Lee et al.(2011)]{lee11} Lee, Y.~S., Beers, T.~C., Allende Prieto, C., et al.\ 2011, \aj, 141, 90 

\bibitem[Lee et al.(2008a)]{lee08a} Lee, Y.~S., Beers, T.~C., Sivarani, T., et al.\ 2008a, \aj, 136, 2022 

\bibitem[Lee et al.(2008b)]{lee08b} Lee, Y.~S., Beers, T.~C., Sivarani, T., et al.\ 2008b, \aj, 136, 2050 




\bibitem[Martin et al.(1988)]{martin88} Martin, G.~A., Fuhr, J.~R., 
\& Wiese, W.~L.\ 1988, New York: American Institute of Physics (AIP) and American Chemical Society, 1988,  

\bibitem[Mashonkina et al.(2007)]{mashonkina07} Mashonkina, L., Korn, A.~J., \& Przybilla, N.\ 2007, \aap, 461, 261 

\bibitem[McWilliam(1997)]{mcwilliam97} McWilliam, A.\ 1997, \araa, 35, 503 

\bibitem[McWilliam et al.(1995)]{mcwilliam95} McWilliam, A., Preston, G.~W., Sneden, C., \& Searle, L.\ 1995, \aj, 109, 2757 

\bibitem[Mel{\'e}ndez et 
al.(2010)]{melendez10} Mel{\'e}ndez, J., Casagrande, L., Ram{\'{\i}}rez, I., Asplund, M., \& Schuster, W.~J.\ 2010, \aap, 515, L3 

\bibitem[Meynet et al.(2006)]{meynet06} Meynet, G., Ekstr\"om, S., \& Maeder, A.\ 2006, \aap, 447, 623

\bibitem[Meynet et al.(2010a)]{meynet10a} Meynet, G., Hirschi, R., Ekstr\"om, S., et al.\ 2010a, in Chemical Abundances in the Universe: Connecting First Stars to Planets, IAU Symposium 265, p. 98

\bibitem[Meynet et al.(2010b)]{meynet10b} Meynet, G., Hirschi, R., Ekstr\"om, S., et al.\ 2010b, \aap, 521

\bibitem[Moity(1983)]{moity83} Moity, J.\ 1983, \aaps, 52, 37 

\bibitem[Nissen \& Schuster(2010)]{nissen10} Nissen, P.~E., \& Schuster, W.~J.\ 2010, \aap, 511, L10 

\bibitem[Nitz et al.(1999)]{nitz99} Nitz, D.~E., Kunau, A.~E., 
Wilson, K.~L., \& Lentz, L.~R.\ 1999, \apjs, 122, 557 

\bibitem[Noguchi et al.(2002)]{noguchi02} Noguchi, K., Aoki, W., Kawanomoto, S., et~al.\ 2002, PASJ, 54, 855

\bibitem[Norris et al.(2012)]{norris12} Norris, J.~E., Bessell, 
M.~S., Yong, D., et al.\ 2012, arXiv:1208.2999 

\bibitem[Norris et al.(2007)]{norris07} Norris, J.~E., Christlieb, N., Korn, A.~J., et al.\ 2007, \apj, 670, 774 

\bibitem[O'Brian et al.(1991)]{obrian91} O'Brian, T.~R., 
Wickliffe, M.~E., Lawler, J.~E., Whaling, W., 
\& Brault, J.~W.\ 1991, Journal of the Optical Society of America B Optical Physics, 8, 1185 



\bibitem[Pickering et al.(2001)]{pickering01} Pickering, J.~C., 
Thorne, A.~P., \& Perez, R.\ 2001, \apjs, 132, 403 

\bibitem[Pinnington et al.(1995)]{pinnington95} Pinnington, E.~H., 
Berends, R.~W., 
\& Lumsden, M.\ 1995, Journal of Physics B Atomic Molecular Physics, 28, 2095 

\bibitem[Press et al.(1992)]{press92} Press, W.~H., Teukolsky, 
S.~A., Vetterling, W.~T., 
\& Flannery, B.~P.\ 1992, Cambridge: University Press, |c1992, 2nd ed.,  


\bibitem[Ryabchikova et al.(1994)]{ryabchikova94} Ryabchikova, T.~A., 
Hill, G.~M., Landstreet, J.~D., Piskunov, N., 
\& Sigut, T.~A.~A.\ 1994, \mnras, 267, 697 

\bibitem[Ryan et al.(1996)]{ryan96} Ryan, S.~G., Norris, J.~E., \& Beers, T.~C.\ 1996, \apj, 471, 254 

\bibitem[Schlegel et al.(1998)]{schlegel98} Schlegel, D., Finkbeiner, D., \& Davis, M. 1998, ApJ, 500, 525

\bibitem[Schnabel et 
al.(2004)]{schnabel04} Schnabel, R., Schultz-Johanning, M., \& Kock, M.\ 2004, \aap, 414, 1169 

\bibitem[Sch{\"o}rck et al.(2009)]{schorck09} Sch{\"o}rck, T., et al.\ 2009, \aap, 507, 817 

\bibitem[Sivarani et al.(2004)]{sivarani04} Sivarani, T., Bonifacio, P., Molaro, P., et al.\ 2004, \aap, 413, 1073 

\bibitem[Skrutskie et al.(2006)]{skrutskie06} Skrutskie, M.~F., Cutri, R.~M., Stiening, R., et al.\ 2006, \aj, 131, 1163 

\bibitem[Smolinski et al.(2011)]{smolinski11} Smolinski, J.~P., Lee, Y.~S., Beers, T.~C., et al.\ 2011, \aj, 141, 89 

\bibitem[Sneden et al.(2008)]{sneden08} Sneden, C., Cowan, J.~J., \& Gallino, R.\ 2008, \araa, 46, 241 


\bibitem[Spite et al.(2012)]{spite12} Spite, M., Andrievsky, S.~M., Spite, F., et al.\ 2012, \aap, 541, A143 

\bibitem[{{Spite} \& {Spite}(1982)}]{spite82}
{Spite}, F., \& {Spite}, M. 1982, \aap, 115, 357

\bibitem[Suda et al.(2004)]{suda04} Suda, T., Aikawa, M., 
Machida, M.~N., Fujimoto, M.~Y., \& Iben, I., Jr.\ 2004, \apj, 611, 476 

\bibitem[Suda et al.(2008)]{suda08} Suda, T., Katsuta, Y., 
Yamada, S., et al.\ 2008, \pasj, 60, 1159 


\bibitem[Tominaga et al.(2007)]{tomonaga07} Tominaga, N., Umeda, H., \& Nomoto, K.\ 2007, \apj, 660, 516

\bibitem[Tsuji(1978)]{tsuji78} Tsuji, T.\ 1978, \aap, 62, 29 

\bibitem[Umeda \& Nomoto(2003)]{umeda03} Umeda, H. \& Nomoto, K.\ 2003, Nature, 422, 871

\bibitem[Umeda \& Nomoto(2005)]{umeda05} Umeda, H. \& Nomoto, K.\ 2005, \apj, 619, 427

\bibitem[Wiese 
\& Martin(1980)]{wiese80} Wiese, W.~L., \& Martin, G.~A.\ 1980, Wavelengths and transition probabilities for atoms and atomic ions: Part 2.~Transition probabilities, NSRDS-NBS Vol.~68.,  

\bibitem[Wiese et al.(1969)]{wiese69} Wiese, W.~L., Smith, 
M.~W., 
\& Miles, B.~M.\ 1969, NSRDS-NBS, Washington, D.C.: US Department of Commerce, National Bureau of  Standards, |c 1969,  


\bibitem[Yong et al.(2012)]{yong12} Yong, D., Norris, J.~E., 
Bessell, M.~S., et al.\ 2012, \apj, in press, arXiv:1208.3003 

\bibitem[Yanny et al.(2009)]{yanny09} Yanny, B., Rockosi, C., Newberg, H.~J., et al.\ 2009, \aj, 137, 4377 

\bibitem[York et al.(2000)]{york00} York, D.~G., Adelman, J., Anderson, J.~E., Jr., et al.\ 2000, \aj, 120, 1579 

\bibitem[Zhang et al.(2009)]{zhang09} Zhang, L., Ishigaki, M., Aoki, W., Zhao, G., \& Chiba, M.\ 2009, \apj, 706, 1095 

\bibitem[Zhao \& Newberg(2006)]{zhao06} Zhao, C., \& Newberg, H.~J.\ 2006, arXiv:astro-ph/0612034


\end{thebibliography}
\end{document}